\documentclass[twocolumn]{aastex631}

\graphicspath{{./}{figures/}{figures/Pagn/}{figures/histogram/}}

\begin{document}

\title{Multiwavelength SED Analysis of X-Ray Selected AGNs at $z=0.2-0.8$ in Stripe 82 Region}

\correspondingauthor{Kenta Setoguchi}
\email{setoguchi@kusastro.kyoto-u.ac.jp}

\author[0000-0001-5353-7635]{Kenta Setoguchi}
\affil{Department of Astronomy, Kyoto University, Kitashirakawa-Oiwake-cho, Sakyo-ku, Kyoto 606-8502, Japan}

\author[0000-0001-7821-6715]{Yoshihiro Ueda}
\affiliation{Department of Astronomy, Kyoto University, Kitashirakawa-Oiwake-cho, Sakyo-ku, Kyoto 606-8502, Japan}

\author[0000-0002-3531-7863]{Yoshiki Toba}
\affiliation{National Astronomical Observatory of Japan, 2-21-1 Osawa, Mitaka, Tokyo 181-8588, Japan}
\affiliation{Academia Sinica Institute of Astronomy and Astrophysics, 11F of Astronomy-Mathematics Building, AS/NTU, No.1, Section 4, Roosevelt Road, Taipei 10617, Taiwan}
\affiliation{Research Center for Space and Cosmic Evolution, Ehime University, 2-5 Bunkyo-cho, Matsuyama, Ehime 790-8577, Japan}

\author[0000-0002-1605-915X]{Junyao Li}
\affiliation{Department of Astronomy, University of Illinois at Urbana-Champaign, Urbana, IL 61801, USA}

\author[0000-0002-0000-6977]{John Silverman}
\affiliation{Kavli Institute for the Physics and Mathematics of the Universe, The University of Tokyo, Kashiwa, Japan 277-8583 (Kavli IPMU, WPI)}
\affiliation{Department of Astronomy, School of Science, The University of Tokyo, 7-3-1 Hongo, Bunkyo, Tokyo 113-0033, Japan}

\author[0000-0001-6653-779X]{Ryosuke Uematsu}
\affiliation{Department of Astronomy, Kyoto University, Kitashirakawa-Oiwake-cho, Sakyo-ku, Kyoto 606-8502, Japan}



\begin{abstract}

We perform a systematic, multiwavelength spectral energy distribution
(SED) analysis of X-ray detected Active Galactic Nuclei (AGNs) at $z=0.2-0.8$ with SDSS
counterparts in the Stripe 82 region, consisting of 60 type-1 and 137 type-2 AGNs covering a 2--10 keV
luminosity range of $41.6 < {\rm log}\ L_{\rm x} < 44.7$. The latest CIGALE code, where dusty polar components
are included, is employed. To obtain reliable host and AGN parameters in type-1 AGNs, we utilize the image decomposed optical SED of host galaxies by Li et al. (2021) based on the Subaru Hyper Suprime-Cam (HSC) images. The mean ratio of black hole masses ($M_{\rm BH}$) and stellar masses ($M_{\rm stellar}$) of our X-ray detected type-1 AGN sample, $\log (M_{\rm BH}/M_{\rm stellar}) = -2.7\pm0.5$, is close to the local relation between black hole and stellar masses, as reported by Li et al. (2021) for SDSS quasars.
This ratio is slightly lower than that found for more luminous ($\log L_{\rm bol} > 45$) type-1 AGNs at $z\sim1.5$.
This can be explained by the AGN-luminosity dependence of $\log (M_{\rm BH}/M_{\rm stellar})$, which little evolves with redshift. 
We confirm the trend that the UV-to-X-ray slope ($\alpha_{\rm OX}$) or
X-ray-to-bolometric correction factor ($\kappa_{2-10}$) increases with
AGN luminosity or Eddington ratio. 
We find that type-1 and type-2 AGNs with the same luminosity ranges share similar host stellar-mass distributions, whereas type-2s tend to show smaller AGN luminosities than type-1s.
This supports the luminosity (or Eddington ratio) dependent unified scheme.

\ifnum0=1
This ratio is also similar to what is found for high luminosity AGNs
(log $L_{\rm x}>44$**) at $z=1.5-3$, implying similar AGN triggering mechanisms
among low luminosity AGNs and high luminosity ones at the redshifts
where their spatial number densities are peaked.
\fi

\end{abstract}

\keywords{Active galaxies (17) --- 
Active galactic nuclei (16) --- Supermassive black holes (1663) }


\section{Introduction} \label{sec:intro}
The cosmological evolution of galaxies and supermassive black holes (SMBHs) in their centers
has been a mainstream of astronomical research. 
Many studies report a tight correlation between the SMBH mass ($M_{\rm{BH}}$) and galactic
classical bulge mass ($M_{\rm{bulge}}$) or the stellar velocity
dispersion in the local universe
($z\sim0$) \citep[e.g.,][]{Magorrian,Ferrarese00,Gebhardt00,Marconi03,Haring,Gultekin,Kormendy}. This
indicates that SMBHs and their host galaxies have co-evolved by
affecting their growths one another \citep[e.g.,][]{Kormendy}.
Active galactic Nuclei (AGNs) represent the processes where
SMBH grows by mass accretion. 
A straightforward approach
to witness the site of SMBH-galaxy coevolution
is to study the properties of host galaxies of
AGNs
(e.g., stellar mass)
and their relations to basic AGN parameters (e.g., 
black hole mass, luminosity, and obscuration).

X-ray observations are a powerful tool to search for AGNs with high completeness, 
thanks to its strong penetrating power against gas and dust, particularly at high photon energies above 2 keV. They also provide 
clean AGN samples because of small contamination by the host galaxy emission. Thus, a useful, widely used technique to tackle the 
issue is multiwavelength spectral energy distribution (SED) analysis of X-ray selected AGNs, which enables one to simultaneously 
constrain both AGN and host properties.

To reveal all processes of SMBH growth, it is important to
study various populations of AGNs. 
In general, AGNs are classified into two types by
optical spectral features: ``type-1'' AGNs, which show
both broad emission lines and narrow ones with typical velocities
of $\sim 2000-20000$ km s$^{-1}$ and $<1000$ km s$^{-1}$,
respectively, and ``type-2'' AGNs, which show
only narrow emission lines. They can also be classified by X-ray
absorption by line-of-sight material: ``unabsorbed''
AGNs with typical hydrogen column
densities of $N_{\rm H} <10^{22}$ cm$^{-2}$ and ``absorbed'' AGNs
with $N_{\rm H} >10^{22}$ cm$^{-2}$. The unified scheme of AGNs \citep{Antonucci93}
explains differences of these AGN properties by the viewing angle with
respect to the dusty torus. 
When intervening in the line of sight, the torus obscures the broad line region and absorbs direct X-ray emission from the hot corona located close to the SMBH.
Generally, the optical and X-ray classifications of an AGN agree with each other
(i.e., type 1 and type 2 AGNs correspond to unabsorbed and absorbed AGNs, respectively).
A fraction of AGNs show mismatched classifications, however;
for instance, \cite{Garcet07} found that 12\% of X-ray selected
AGNs had intrinsically differing X-ray and optical classifications.
The origins of the mismatches are still under debate (e.g., see \citealt{Ogawa21}). 
Verification of the AGN unified scheme has been one of the fundamental issues in understanding AGN phenomena.

\subsection{Goals of This Work}

The main focus of this paper is two-fold. The first
  immediate objective is to establish the relation of SMBH mass and host stellar mass of X-ray selected AGNs as a function of AGN luminosity and redshift.
For this purpose, we systematically analyze the SEDs of X-ray selected AGNs at redshifts $z<0.8$ in the SDSS Stripe 82 region (see Section~1.2), one of the best-studied multiwavelength fields covering a large area (31.3 deg$^2$), 
and combine the results with those obtained at $z\sim1.5$ from
the Subaru/XMM-Newton Deep Field (SXDF; \citealt{Furusawa08,Ueda08,Akiyama,Setoguchi21}),
a deeper multiwavelength field covering an area of $\sim1$ deg$^2$.
The ultimate goal of this study is to reveal the origin of the cosmological evolution of the AGN luminosity function,
  as explained in the next subsubsection.
  The second objective is to test if the ``AGN unified scheme'' is valid in X-ray AGNs.
  The unified scheme assumes that all AGNs belong to intrinsically the same population, meaning that their host galaxies should be the same among different AGN types. This can be checked by comparing the host properties of type 1 and type 2 AGNs with similar AGN properties.

\subsubsection{Relation between SMBH Mass and Stellar Mass}

Past X-ray surveys of AGNs have revealed that the comoving spatial
number density of lower luminosity AGNs shows a peak at lower
redshifts than that of higher luminosity ones. This is often referred to as a ``cosmic downsizing'' phenomenon.
A similar downsizing trend has been also found for
galaxy evolution (e.g., \citealt{Cowie96,Kodama04,Fontanot09}), supporting an SMBH-galaxy coevolution
scenario. The origin of SMBH downsizing is still under debate
(e.g., \citealt{Fanidakis12,Draper12,Enoki14,Shirakata19}). Some authors (e.g.,\citealt{Draper12}) suggest that there are two channels of SMBH growth, via major galaxy mergers and secular processes within a single galaxy, which are responsible for activating more luminous AGNs at higher redshifts and less luminous ones at lower redshifts, respectively. 
Thus, it is of great interest to investigate host/AGN properties of low
luminosity AGNs (log $L_{\rm x}<43.5$) at $z<1$ where their number density is
peaked, and to compare them with those of higher luminosity AGNs at $z>1$.


The relation between SMBH mass and host stellar mass gives us 
an important clue
to test whether the underlying mechanisms of coevolution are
the same or not among AGNs at different redshift and luminosity ranges.
Type-1 AGNs are ideal objects for this study 
because $M_{\rm{BH}}$ can be
directly determined through the measurements of broad-line widths and
continuum luminosities by single-epoch optical spectroscopy
(e.g., \citealt{Vestergaard06,Jahnke09,Merloni10,Assef11,Rakshit20}).
Since it is difficult to spatially separate the bulge and disk
components in the distant universe, the relation between $M_{\rm{BH}}$
and total stellar mass ($M_{\rm{stellar}}$) has been intensively
investigated to discuss the coevolution in broad-line AGNs (e.g.,
\citealt{Shields03,Matsuoka15,Sun15,Reines15,Yue18,Suh20,Ishino20,Ding20,Mountrichas23}).
It has been an issue to reliably estimate the total stellar mass or SMBH mass
in a type-1 AGN, however, because
it is often difficult to
separate the contributions from the host galaxy and nucleus
in the SED analysis.
In a very luminous AGN, it is challenging to accurately extract the spectrum of the host galaxy because the AGN dominates the
infrared-optical-UV emission, which could cause a large uncertainty in $M_{\rm{stellar}}$ (e.g., \citealt{Toba18,Toba22}). In a lower luminosity AGN, the AGN spectrum can be significantly affected by contamination from the host galaxy lights, making it difficult to accurately estimate AGN parameters (e.g., $M_{\rm{BH}}$, bolometric luminosity ($L_{\rm bol}$),
UV/optical-to-X-ray spectral index ($\alpha_{\rm ox}$)
\footnote{It is defined as $\alpha_{\rm OX}=-0.3838 \log L_{\rm 2500 \AA}/L_{\rm 2keV}$, where $\log L_{\rm 2500 \AA}$ and $L_{\rm 2keV}$ are intrinsic (extinction-corrected) luminosities at rest-frame 2500$\AA$ and 2 keV, respectively.}
).  

To overcome these problems, high-resolution optical images that allow one to
{\it spatially} decomposite the nucleus and host components are useful.
\cite{Ishino20} investigated 862 type-1 Sloan Digital Sky Survey
(SDSS) quasars and their host galaxies properties at $z<1$ using
Subaru Hyper-Suprime Cam (HSC; \citealt{Miyazaki18}) data (HSC Subaru Strategic Program
(SSP); \citep{Aihara18}) and one-dimensional profile fitting.
\cite{Li21a} carried out a two-dimensional profile fitting to 4887
type-1 SDSS quasars and measured the host galaxy flux, effective
radius ($r_{\rm e}$), and S\'{e}rsic index ($n$). Then, by performing SED
fitting to the optical photometries of the host galaxies, \cite{Li21a}
investigated the evolution of $M_{\rm BH}-M_{\rm stellar}$ relation of type-1
SDSS quasars at $0.2<z<0.8$. In this paper, we basically follow their approach to
separate the contributions from the host galaxy and AGN in SED analysis, which enables us to best
estimate both $M_{\rm BH}$ and $M_{\rm stellar}$ for our X-ray selected type-1 AGN sample
(Section \ref{sec:analysis}).

\ifnum0=1
X-ray parameter gives us a hint to reveal the SMBH evolution scenario.
For example, the UV/optical-to-X-ray spectral index $\alpha_{\rm OX}$, which represents the ratio of UV to X-ray luminosity ($\alpha_{\rm OX}=-0.3838\ \rm{log}\  L_{\rm 2500 \AA}/L_{\rm 2keV}$), is determined to explore the accretion mechanism onto SMBHs (e.g., \citealt{Lusso10,Laurenti22,Liu21,Martocchia17,Zappacosta20,Pu20}).
\cite{Hickox09} explored radio, infrared(IR), and X-ray selected AGN properties at $z=0.25-0.8$ and discovered that radio-selected AGNs have galaxies classified as red sequence galaxies and low $\lambda_{\rm Edd}$ ($<10^{-3}$), the hosts of X-ray selected AGNs show the green valley and IR-selected AGNs have bluer and less luminous galaxies than X-ray selected AGNs. Thus, in different populations, SMBHs and hosts may experience different evolution pathways and we need to evaluate differences in their properties.
\fi

\subsubsection{Testing AGN Unified Scheme}

X-ray surveys performed at energies above 2 keV also provide a large
number of obscured (type-2) AGNs, which are a dominant AGN population
at low to moderate luminosity ranges \citep{Toba13,Ueda14}.
As mentioned earlier, 
comparing the basic properties of host galaxies among type-1 and type-2 AGNs
is always important to test the AGN unified scheme \citep{Antonucci93} and
possible AGN-type dependence on the environments.
\cite{Bornancini18} find no significant difference of $M_{\rm stellar}$ distribution in type-1 and type-2 AGNs at $z=0.3-1.1$, consistent with the prediction
from the unified scheme. 
By contrast, \cite{Zhuang20} concluded that type-2 AGNs show stronger star formation activities than type-1 AGNs at $z<0.3$ regardless of their $M_{\rm stellar}$, $\lambda_{\rm Edd}$, and molecular gas mass.
\cite{Zou19} and \cite{Mountrichas21} suggest that type-1 and type-2 AGNs reside in hosts with similar star formation rates but with smaller and larger
stellar masses, respectively.

\subsection{Survey Field: Stripe 82 Region}

The SDSS Stripe 82 region is one of the most intensively studied,
wide-area multiwavelength survey fields, on which medium-depth X-ray surveys with XMM-Newton and Chandra have been
performed. \cite{LaMassa16} and \cite{Ananna17} present X-ray source
catalogs in this field (Stripe 82X) with multiwavelength photometries
over the radio, infrared, optical, UV, and X-ray bands. The catalog
covers an area of 31.3 deg$^2$ area. Optical spectroscopic
completeness of the total X-ray sources is 43\%. 
The Stripe 82X catalog is useful not only for studies of X-ray selected AGNs \citep{LaMassa16b,LaMassa17,LaMassa19} 
  but also for other AGN populations such as mid-infrared selected AGNs \citep{Glikman18}.
The black hole masses
of type-1 AGNs were available in the catalogs of \cite{Paris18} and
\cite{Rakshit20}, which were calculated from the line widths of Mg II,
H$\beta$, or C IV lines and monochromatic luminosities at 5100 \AA
($L_{\lambda 5100}$) by analyzing the SDSS spectra. The large X-ray
sources catalog with multiwavelength datasets, reliably decomposed
host galaxy SEDs by the HSC images at $z=0.2-0.8$, and $M_{\rm BH}$ in
type-1 AGNs offer us an ideal opportunity to investigate the AGN and
host connection at low redshifts ranges where low to moderate AGNs
have number-density peaks.

\subsection{Outline of This Paper}

In this paper, we perform a systematic multiwavelength SED analysis of X-ray AGNs matched to the SDSS spectroscopic catalog in the Stripe 82
region. We utilize the latest {\tt CIGALE} code \citep{Yang22} where polar dusty
components are included in the AGN template. Thus, our sample contains X-ray detected optical type-1 AGNs whose host photometries are decomposed from the HSC images by \cite{Li21a} and X-ray detected type
2 AGNs. For type-1 AGNs, we fix the host galaxy parameters as
determined by \cite{Li21a} to reliably separate the AGN
emission. Below are the outlines of this paper. In Section
\ref{sec:analysis}, we provide a detailed description of the sample
selection and the technique utilized for SED fitting. In Section
\ref{sec:resultsanddiscussion}, we statistically study the relations
among AGN parameters and host stellar mass and compare them between
type-1 and type-2 AGNs.  We also discuss the multiwavelength SED of
AGNs in terms of $\alpha_{\rm OX}$ or 2--10 keV to bolometric
correction factor ($\kappa_{2-10}$). Section \ref{sec:conclusion}
summarizes the conclusions drawn from our research. We adopt
cosmological parameters of $H_0$=70.4 km s$^{-1}$ Mpc$^{-1}$,
$\Omega_M=0.272$, and $\Omega_{\Lambda}=0.728$ (the Wilkinson Microwave Anisotropy Probe 7 cosmology: \citealt{Komatsu11}).


\section{Data and Analysis} \label{sec:analysis}

\subsection{Sample Selection} \label{subsec:selection}

\cite{Li21a} performed a 2-dimensional image decomposition analysis
for 4887 host galaxies of SDSS-detected type-1 quasars at $z=0.2-0.8$,
including those in the Stripe-82 region. The
  point-source component (corresponding to the quasar) and the host
  galaxy component were fitted with a model consisting of the point spread function (PSF)
  model and a 2D S\'{e}rsic profile, respectively.
They determined the host galaxy parameters (e.g., host galaxy fluxes of HSC g, r, i, z, and y (hereinafter called \textit{grizy}) filters,
effective radius $r_{\rm e}$, S\'{e}rsic index $n$, and ellipticity
$\epsilon$). 
2424 out of 4887 objects are classified as a final sample
whose selection criteria are defined by \cite{Li21a}. 
The selection criteria relevant to our study are as follows: 
\begin{enumerate}
    \item $z<0.8$
    \item The derived $M_{\rm stellar}$ meets the stellar mass cut criteria: log $M_{\rm stellar,cut}<$ log $M_{\rm stellar}<$ 11.5, where log $M_{\rm stellar,cut}$ is 9.3 ($z=0.2-0.4$), 9.8 ($z=0.4-0.6$), and 10.3 ($z=0.6-0.8$).
    \item The reduced $\chi^2$ of the SED fitting is smaller than 10.
\end{enumerate}

\cite{Li21a} cataloged 371 optical type-1 AGNs at $z=0.2-0.8$ within the Stripe 82X region.
Among them, we selected 111 objects
detected with XMM-Newton and/or Chandra 
that have
SDSS DR14
spectroscopic redshifts and multiwavelength counterparts in
\cite{Ananna17} and \cite{LaMassa16}.
To guarantee the reliability in $M_{\rm BH}$ and counterpart matching,
we chose 81 objects with $M_{\rm BH}$ Quality Flag (QF)=0 in the SDSS DR14Q catalog and QF=1--2 in the Stripe 82X catalog.
The former QF condition ensures the quality of host galaxy decomposition by
a principal component analysis (PCA;\citealt{Yip04a,Yip04b}) in estimating the
continuum luminosities and widths of broad emission lines. The latter 
corresponds to reliable multiwavelength identification
by excluding the cases where different counterparts are found in multiple bands with comparable likelihood ratios
or there is a counterpart in only one band
(see \citealt{Rakshit20,Ananna17}). 

For correcting the systematic biases, \cite{Li21a} calibrated host galaxy
fluxes of \textit{grizy} bands using simulated galaxy and AGN datasets.
To minimize any possible calibration uncertainties, we selected the
objects with small differences between the fitted and calibrated fluxes in the HSC $i$-band. The criterion we adopt was $|F_{\rm fit}-F_{\rm cal}|/F_{\rm cal}<0.3$, where $F_{\rm fit}$ and $F_{\rm cal}$ are the fitted flux (before calibration) and calibrated flux (after calibration), respectively. 
\cite{Li21a} decomposed the HSC images to derive the host galaxy parameters including $F_{\rm fit}$. However, galaxy structural measurements can have significant biases, like underestimating the size of large galaxies, due to
  various effects
  such as PSF blurring, limited signal-to-noise ratios, and reduced surface brightness. To address these biases, \cite{Li21a} employed two calibration methods: inserting model galaxies into empty areas of real HSC images and adding unresolved quasars using model PSFs to real HSC images of galaxies in the CANDELS field. After these calibrations, $F_{\rm cal}$ is calculated.
Finally, 66 objects whose fluxes are decomposed in all {\textit{grizy}} bands were chosen.

\ifnum0=1
\cite{Rakshit20} estimated the black hole masses using the broad-line
FWHMs (Mg II, C IV, or H$\beta$) and continuum luminosities at 5100
\AA\ for SDSS type-1 quasars in the SDSS DR14Q catalog \citep{Paris18}, including our type-1 AGN sample.
We
calculate the Eddington ratio as $\lambda_{\rm Edd} = L_{\rm
  bol}$/$L_{\rm Edd}$, where $L_{\rm Edd} = 1.3\times10^{38} M_{\rm
  BH}/M_\odot$.  The black hole masses of our type-1 AGNs span a range
of $\log M_{\rm BH}/M_\odot = 6.95 \sim9.34 $ (with a median of 8.05).
\fi


We also selected X-ray selected, optical type-2 AGNs at $z=0.2-0.8$
from the Stripe 82X catalog. In addition to the QF criterion regarding cross-matching, we imposed the selection criteria as follows:
\begin{enumerate}
\setlength{\parskip}{1mm}
    \item No counterparts are found in SDSS DR14Q \citep{Paris18} within 1\arcsec of the position.
    \item No broad lines are observed in the optical spectrum. (In the SDSS catalog, if emission lines are detected at $>10\sigma$ level with a width of $> 200$ km s$^{-1}$ at 5$\sigma$ level, objects are classified as ``BROADLINE'' in their subclasses.)
    \item Spectroscopic redshifts are determined with no warning flags (i.e., ZWARNING=0 in the SDSS catalog).
    \item Spectral type are classified as galaxies by the PCA in the SDSS optical spectroscopy pipeline
      (see Section 3.2 of \cite{Paris18} for more details).
    \item The ``morphology'' column in the Stripe 82X catalog is not assigned to type-1 AGNs or QSOs. In \cite{Ananna17}, firstly an object is classified as a point-like or extended source on the basis of optical and NIR morphology obtained by \citet{Fliri16}, \citet{Jiang14}, and/or \citet{McMahon13}. Secondly, according to the photometries and the image classification (point-like or extended), a limited set of SED templates are selected from those for stars, elliptical/spiral/starburst galaxies, type-1/type-2 AGNs, and QSOs (\citealt{Ilbert09,Salvato09,Hsu14}), 
      and its ``morphology'' is finally determined via template fitting (see Sections 3.1, 3.3, and Table 5 in \citealt{Ananna17} for details). 
\end{enumerate}
This selection left 158 type-2 AGNs.

\subsection{SED Fitting with {\tt CIGALE}}\label{sec:CIGALE}

\begin{table}
\renewcommand{\thetable}{\arabic{table}}
\centering
\caption{
Grid of parameters
used for the host SED fitting in type-1 AGNs with 
{\tt CIGALE}.}
\label{T1}
\begin{tabular}{lc}
\tablewidth{0pt}
\hline
\hline
Parameter & Value\\
\hline
\multicolumn2c{Delayed SFH}\\
\hline
$\tau_{\rm main}$ [Myr] & 40, 63, 100, 158, 251, 398,\\
& 631, 1000, 1584, 2512, 3981, 6309, \\
& 10000, 15848, 25118 \\
$\tau_{\rm burst}$ [Myr] & 50 \\
$f_{\rm burst}$ & 0.0 \\
age [Myr] & 40, 63, 100, 158, 251, 398,\\
& 631, 1000, 1584, 2512,\\
& 3981, 6309, 10000 \\
\hline
\multicolumn2c{Single stellar population \citep{Bruzual}}\\
\hline
IMF & \cite{Chabrier} \\
Metallicity & 0.02 \\
\hline
\multicolumn2c{Dust attenuation modified starburst}\\
\hline
$E(B-V)_{\rm lines}$ & 0.01, 0.05, 0.1, 0.15, 0.2\\
& 0.25, 0.3, 0.35, 0.4, 0.45\\
& 0.5, 0.55, 0.6, 0.65, 0.7, \\
& 0.75, 0.8, 0.85, 0.9, 0.95, 1.0 \\
${\rm Extinction\ law}$ & 1, 3 \\
\hline
\end{tabular}
\end{table}

\begin{figure*}[ht!]
\gridline{\fig{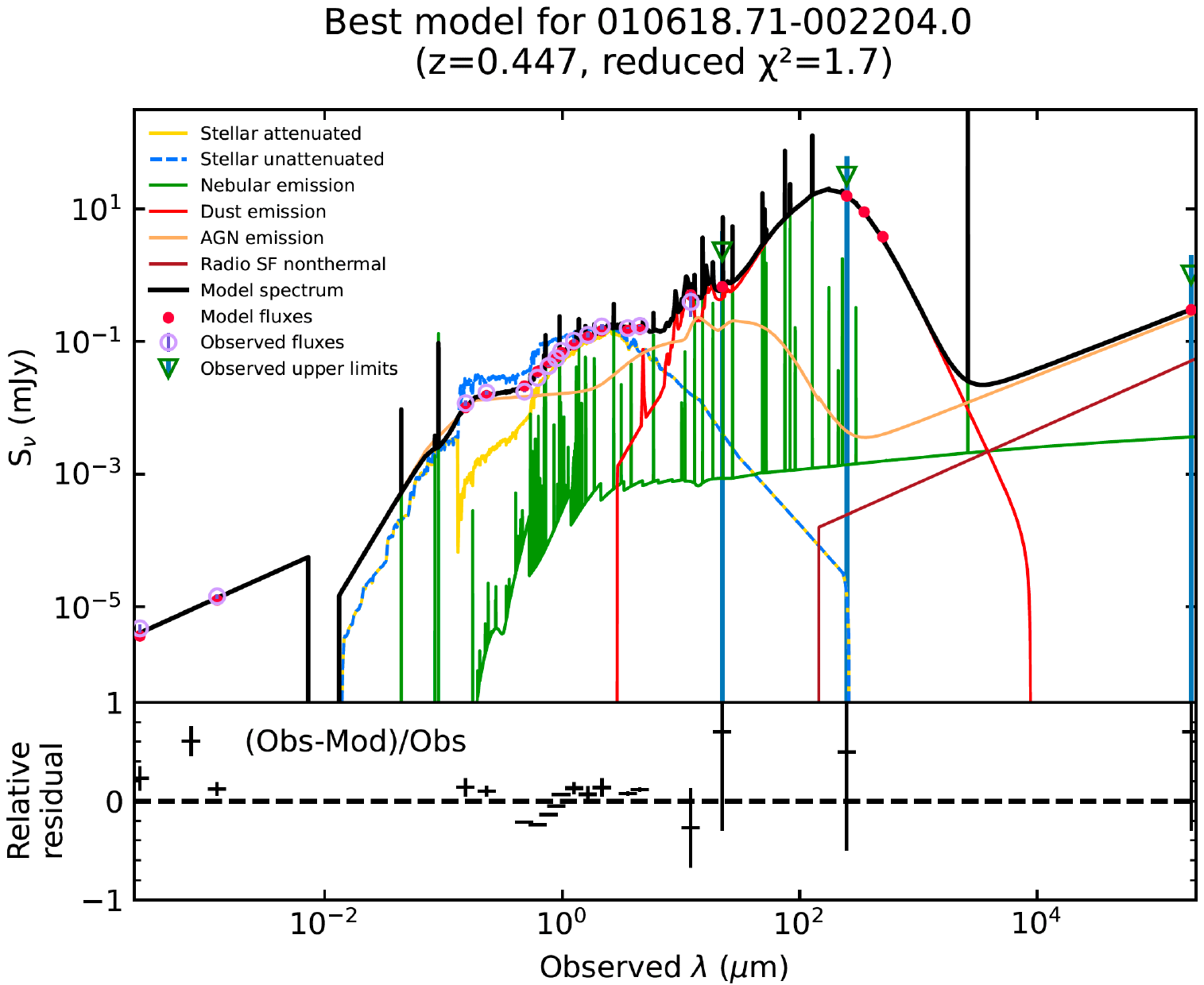}{0.8\textwidth}{(a)}}
\gridline{\fig{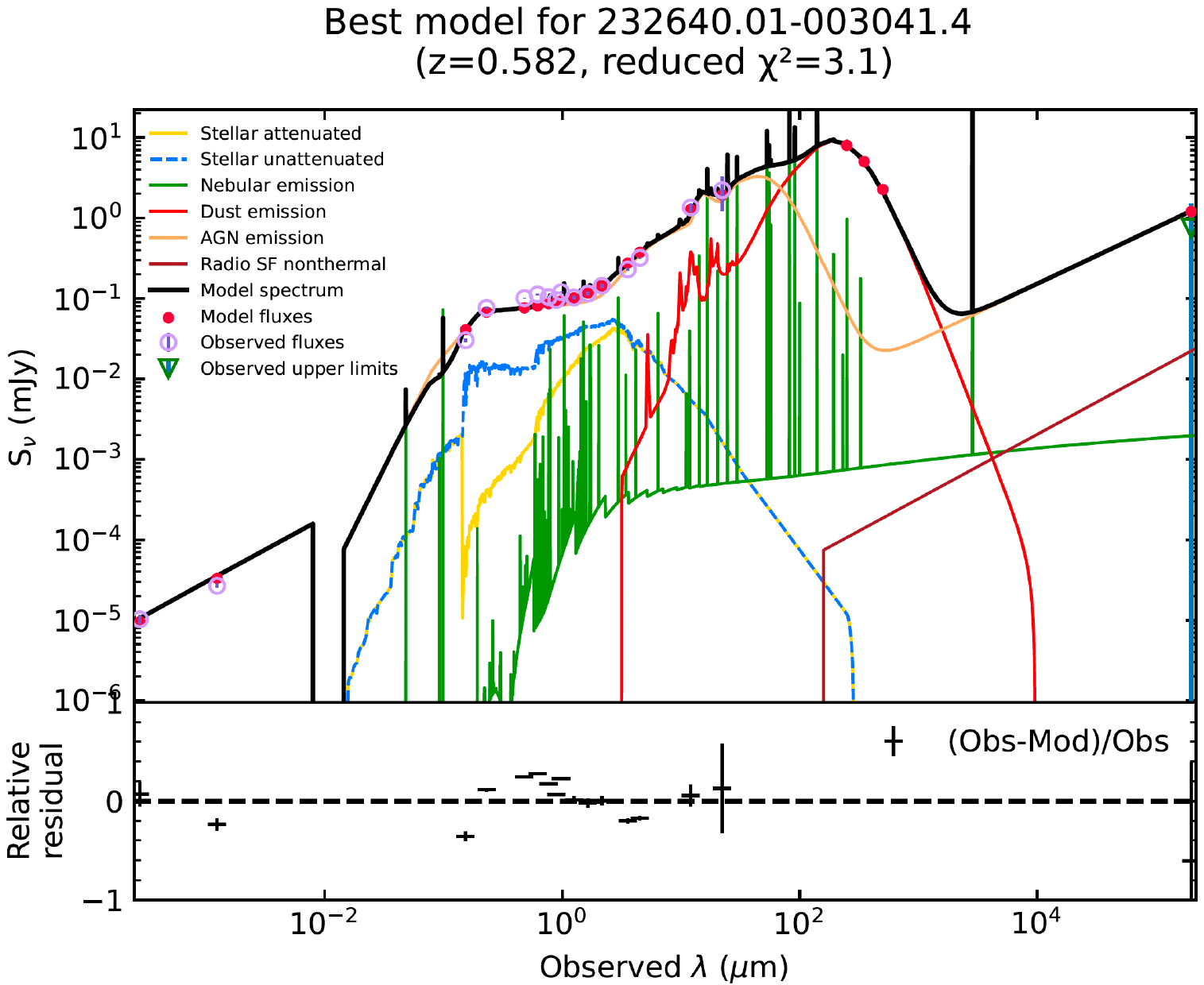}{0.8\textwidth}{(b)}}
 \caption{(a), (b) Examples of the total (host+AGN) SED fitting for type-1 AGNs dominated by host galaxy emission (left; 010618.71-002204.0) and by AGN emission (right; 232640.01-003041.4) in the optical bands. The black solid lines show the best fit SEDs.}
 \label{fig:cigale}
\end{figure*}

  We performed X-ray to radio SED modeling for a sample of 224 AGNs (type-1: 66 sources, type-2: 158 sources) utilizing a new version of \texttt{CIGALE} (Code Investigating GALaxy Emission version 2022.0; \citealt{Burgarella,Noll,Boquien19,yang2020x,Yang22}). \texttt{CIGALE} assumes the energy balance between UV/optical absorption and far-infrared emission, which enables us to model a multiwavelength 
  SED in a self-consistent way. This code is designed to calculate the likelihoods of all the models on a user-defined grid and return the likelihood-weighted mean of a physical quantity by Bayesian estimation. 
It is noteworthy that {\tt CIGALE} is able to take into account upper limits in the photometric data,
using the method of \cite{Sawicki12} as described in Section 4.3 of \cite{Boquien19}.

\texttt{CIGALE} offers several options for the SED templates of each component. 
In this work, we adopted the same module selection as in \cite{Toba21} and \cite{Setoguchi21}.
We employed a delayed star formation history (SFH) model, where $\tau_{\rm{main}}$ represents the e-folding time of the main stellar population. The simple stellar population (SSP) was modeled by the stellar templates of \cite{Bruzual} and the Chabrier initial mass function \citep[IMF;][] {Chabrier}. We utilized the default template by \cite{Inoue} for the nebular emission. In order to consider the dust attenuation
to the stellar components,
we adopted the extinction curve by \cite{Calzetti} and \cite{Leitherer02}, which is characterized by the color excess $(E(B-V)_*)$. 
The reprocessed IR dust emission of UV/optical stellar radiation is modeled with the templates by \cite{Dale}.
For the optical to infrared emission of an AGN, we used the two-phase torus model named {\tt SKIRTOR} \citep{Stalevski16}, incorporating polar dust emission with a single modified blackbody.
This model takes into account extinction and re-emission of the direct AGN component by the torus and polar dust.
In the following subsection, we show the details of the SED modeling of type-1 and type-2 AGNs, separately.

\subsubsection{Type-1 AGN Sample}
As described in Section \ref{sec:intro}, in a type-1 AGN, the host component can be largely contaminated by
  the AGN component in the optical band,
  because the nuclear emission is not obscured by the dusty torus and hence is much brighter than in a type-2 AGN.
To obtain reliable host and AGN parameters by separating the two components in type-1 AGNs, we analyzed their SEDs in two steps: (1) 
optical SEDs of host galaxies decomposed by \cite{Li21a}
and (2) infrared, optical, UV, and X-ray SED of the total
emission including host and AGN components.

In the first step, we analyzed the optical 
host SEDs based on the fitted fluxes in \citet{Li21a}, 
utilizing the same SED modules as adopted by \citet{Li21a}.
In this stage, dust re-emission components were ignored.
Table \ref{T1} details the free parameter ranges of the host SED analysis. 
We obtain reasonable fit with reduced $\chi^2$ values less than 10 
for all 66 objects. 
The obtained stellar masses are confirmed
to be fully consistent with those reported in \cite{Li21a}.

\begin{table}
\renewcommand{\thetable}{\arabic{table}}
\centering
\caption{Grid of parameters used for the total (host+AGN) SED fitting with 
{\tt CIGALE}.}
\label{T2}
\begin{tabular}{lc}
\tablewidth{0pt}
\hline
\hline
Parameter & Value\\
\hline
\multicolumn2c{Delayed SFH}\\
\hline
$\tau_{\rm main}$ [Myr] & The value of the host SED \\
& (40, 63, 100, 158, 251, 398,\\
& 631, 1000, 1584, 2512, 3981, \\
& 6309, 10000, 15848, 25118) \\
age [Myr] & The value of the host SED \\
& (40, 63, 100, 158, 251, 398, \\
& 631, 1000, 1584, 2512, \\
& 3981, 6309, 10000) \\
\hline
\hline
\multicolumn2c{Dust attenuation modified starburst}\\
\hline
$E(B-V)_{\rm lines}$ & The value of the host SED \\
& (0.01, 0.05, 0.1, 0.15, 0.2, \\
& 0.25, 0.3, 0.35, 0.4, 0.45, \\
& 0.5, 0.55, 0.6, 0.65, 0.7, \\
& 0.75, 0.8, 0.85, 0.9, 0.95, 1.0) \\
${\rm Extinction\ law}$ & 1, 3 \\
\hline

\multicolumn2c{Dust emission \citep{Dale}}\\
\hline
IR power-law slope ($\alpha_{\rm dust}$) & 0.0625, 1.0000, 1.5000, 
2.0000, \\
& 2.5000, 3.0000, 4.0000 \\
\hline
\multicolumn2c{AGN emission \citep{Stalevski16}}\\
\hline
$\tau_{\rm 9.7}$ 			& 	3, 5, 9		\\
$p$							&	0.0, 1.0, 1.5	\\
$q$							&	0.0, 1.0, 1.5	\\
$\Delta$					&	10, 30, 50\arcdeg	\\
$R_{\rm max}/R_{\rm min}$ 	& 	30 			\\
$\theta$					&	30\arcdeg	\\
$f_{\rm AGN}$ 				& 	0.1, 0.3, 0.5, 0.7, 0.9, 0.99\\
\hline
\multicolumn2c{AGN polar dust emission \citep{yang2020x}}\\
\hline
$E(B-V)$                    & 	0.0, 0.05, 0.1, 0.15, 0.2, 0.3, 0.4	\\
$T_{\rm dust}^{\rm ploar}$	&	100.0, 500.0, 1000.0	\\
\hline
\multicolumn2c{Xray emission}\\
\hline
$\alpha_{\rm ox}$			&	-1.9, -1.8, -1.7, -1.6, -1.5\\
& -1.4, -1.3, -1.2, -1.1	\\
max dev $\alpha_{\rm ox}$	&	0.0 \\
\hline
\end{tabular}
\end{table}

In the second step, 
we performed a multi-component SED fitting to 19 photometries in the radio(VLA), far-IR (Herschel/SPIRE), mid-IR
(Spitzer/IRAC and WISE), near-IR (VISTA), optical (Subaru),
ultraviolet (GALEX), and X-ray (XMM-Newton or/and Chandra) bands for
each object (see \citealt{Ananna17} and \citealt{LaMassa16} for
details). The photometries except for the optical bands
were taken from the Stripe 82X catalog.
To obtain the non-decomposed optical photometries from the Subaru data, we performed a
nearest-neighbor matching between the Subaru/HSC images and Stripe
82X catalog within 1\arcsec.
We assigned an upper flux limit if
the object was observed but was not detected in that band 
(see Section \ref{subsubsec:mirfir} for the mid- and far-infrared photometries).
We corrected the X-ray fluxes for absorption, if any, according to the recipe described in Section \ref{subsec:Lx}.

Here we considered dust re-emission and AGN
components, which were not considered in the first step.
As mentioned earlier, we utilized 
the {\tt SKIRTOR} model for the AGN emission, which has
seven parameters: torus optical depth at 9.7 $\micron$ ($\tau_{\rm 9.7}$), torus density radial parameter ($p$), torus density angular parameter ($q$), the angle between the equatorial plane and the edge of the torus ($\Delta$), the maximum to minimum radii ratio of the torus ($R_{\rm max}/R_{\rm min}$), the viewing angle ($\theta$), and the fraction of AGN contribution to the total IR luminosity ($f_{\rm AGN}$).
We fixed $R_{\rm max}/R_{\rm min}$ and $\theta$ at values typically observed in type-1 AGNs,
following \cite{yang2020x}.
When we performed total-SED analysis for the type-1 AGNs,
 the normalization factor in {\tt CIGALE} was set to be unity
 to fix the host-galaxy component to that obtained from the host-SED analysis.
Table \ref{T2} summarizes the free parameters in the total SED model.
We obtain reasonable fits with reduced $\chi^2<10$ (the same threshold as adopted in
\cite{Setoguchi21}) for 60 out of the 66 objects.
The worse fits for the remaining objects could be
caused by
our too-simplified SED models and/or time variability in AGN emission.
Hereafter we refer to this sample as the ``(X-ray detected) type-1 AGN sample''.
%
The AGN emission module, SKIRTOR, enables us to estimate ``agn.accretion power'', which is the intrinsic AGN disk luminosity averaged over all directions. In this work, we adopt this parameter as $L_{\rm bol}$ in both type-1 and type-2 AGNs.
Table \ref{table:parameter_catalog} lists the best-fit parameters of $M_{\rm stellar}$ and $L_{\rm bol}$ for each
object derived from the SED fitting together with the basic source information.
Examples of the total SED fitting in type-1 AGNs are represented in Figure \ref{fig:cigale}.

\subsubsection{Type-2 AGNs}

For type-2 AGNs, we only perform the second step (i.e., analysis of
the IR-to-X-ray SED of the total emission) without fixing
the host parameters. This is because, due to the extinction of the AGN component, the optical SED is dominated by the host galaxy and hence the stellar mass is reliably constrained.
The range of the free parameters is listed in Table~\ref{T3}. We
confirmed that the SEDs of 137 objects out of 158
are reproduced with reduced $\chi^2<10$ and satisfy the stellar mass
cut criteria imposed by \cite{Li21a} (see Section \ref{subsec:selection}).
Hereafter we refer to these 137 AGNs as the ``type-2 AGN'' sample.
The estimated $L_{\rm bol}$ and $M_{\rm stellar}$ are listed in Table~\ref{table:parameter_catalog_type2}.

\subsubsection{Notes on Mid- and Far-Infrared Data}
\label{subsubsec:mirfir}

In this subsubsection, we discuss the mid- to far-infrared data quality of our sample. The mid-infrared photometries of our sample are given by the all-sky WISE Mission (AllWISE), the Spitzer-HETDEX Exploratory Large Area Survey (SHELA; \citealt{Papovich16}), and the Spitzer IRAC Equatorial Survey (SpIES; \citealt{Timlin16}). While AllWISE covers the entire Stripe 82 region, both SHELA and SpIES only cover it partially. Most of our objects (222/224) are detected in at least one mid-infrared band. This enables us to constrain the mid-infrared SEDs of the AGNs and to reliably estimate the bolometric AGN luminosities. 
The far-infrared photometries are extracted from the Herschel Stripe 82 Survey (HerS; \citealt{Viero14}), which also covers only a portion of the Stripe 82 region. From our sample, 194/224 lie within the coverage, and 7/194 are detected in at least one band. This relatively low detection rate can be attributed to the limited depth of HerS. The flux limit of HerS is 31 mJy at 250 $\mu$m (3$\sigma$), which corresponds to $\mathrm{SFR}=15\text{--}115\, M_{\odot}\, \mathrm{yr}^{-1}$ at $z=0.2\text{--}0.8$.\footnote{Here we convert the 250 $\mu$m flux density to the SFR using a typical SED of a star-forming galaxy. First, we convert the 250 $\mu$m flux density to the dust luminosity assuming a single optically-thin graybody with an emissivity index of 2.0 and a temperature of 30 K \citep{Drew22}. Then, we convert the dust luminosity to the SFR following \citet{Kennicutt98}, where we multiply 0.63 to convert the Salpeter IMF to the Chabrier IMF.} This may provide the upper limit of the SFRs of our sample. Here we note that the SFRs of our sample are not well-constrained due to the limited far-infrared photometries. However, the stellar masses are reliably constrained from the optical to near-infrared SEDs \citep{Conroy13}.

\startlongtable
\begin{deluxetable*}{cccccc}
\tablecaption{Summary of properties of X-ray detected type-1 AGNs and hosts}
\tablewidth{0pt}
\tablehead{
\colhead{ID (SDSS DR14Q)} & \colhead{SpecObjID} & \colhead{redshift} & \colhead{log $L_{\rm bol}\ ({\rm erg\ s}^{-1})$} & \colhead{log $M_{\rm BH}\ (M_\odot)$} & \colhead{log $M_{\rm stellar} \ (M_\odot)$} }
\decimalcolnumbers
\startdata 
\hline
001010.03+005126.6 & 438068002609457152 & 0.387 & 44.53 & 7.42 $\pm$ 0.26 & 10.89 $\pm$ 0.17 \\
001103.18+005927.2 & 772532854855854080 & 0.486  & 44.53 & 7.80 $\pm$ 0.10 & 10.29 $\pm$ 0.18 \\
010230.03--003206.8 & 444737365064312832 & 0.343 & 44.47 & 8.64 $\pm$ 0.05 & 10.95 $\pm$ 0.16 \\
010235.46--002624.5 & 9894469527709655040 & 0.43 & 43.83 & 8.03 $\pm$ 0.19 & 10.26 $\pm$ 0.18 \\
010524.21+000901.2 & 9895675417053863936 & 0.418 & 44.33 & 7.06 $\pm$ 0.58 & 9.97 $\pm$ 0.17 \\
010618.71--002204.0 & 780256094961297408 & 0.447 & 44.20 & 7.39 $\pm$ 0.39 & 10.48 $\pm$ 0.17 \\
010712.69+000348.3 & 9894418400418963456 & 0.531 & 44.31 & 8.61 $\pm$ 0.09 & 10.69 $\pm$ 0.17 \\
010737.01--001911.6 & 447054311126493184 & 0.738 & 45.74 & 8.20 $\pm$ 0.15 & 10.96 $\pm$ 0.20 \\
010739.79+002423.0 & 4758265578795278336 & 0.518 & 44.45 & 8.28 $\pm$ 0.25 & 10.52 $\pm$ 0.18 \\
010811.00+000932.9 & 8865536649473269760 & 0.584 & 44.46 & 7.60 $\pm$ 0.20 & 10.24 $\pm$ 0.22 \\
010823.65+001249.4 & 9895691359972466688 & 0.733 & 44.36 & 7.93 $\pm$ 0.22 & 10.40 $\pm$ 0.23 \\
011130.50+001427.9 & 9895746335553855488 & 0.432 & 44.11 & 7.37 $\pm$ 0.17 & 10.69 $\pm$ 0.17 \\
011159.58--001614.2 & 782574139478140928 & 0.458 & 44.31 & 7.36 $\pm$ 0.26 & 10.34 $\pm$ 0.18 \\
011200.57--002336.9 & 447021050899752960 & 0.747 & 45.03 & 8.67 $\pm$ 0.22 & 10.83 $\pm$ 0.23 \\
011214.01--002615.3 & 4206379181871702016 & 0.44 & 44.13 & 7.52 $\pm$ 0.34 & 9.97 $\pm$ 0.16 \\
\hline 
\enddata
\tablecomments{(1) Unique identifier in the SDSS DR14Q catalog. (2) Unique spectrum identifier in SDSS DR14. (3) Redshift from the SDSS spectra. (4) Logarithmic bolometric AGN luminosity derived by CIGALE, for which we adopt the output parameter ``agn.accretion power''. (5) Newly estimated logarithmic black hole mass and its 1$\sigma$ error. (6) Logarithmic stellar mass derived by CIGALE and its 1$\sigma$ error. 
(This table is available in its entirety in machine-readable form.)}
\label{table:parameter_catalog_type2}
\end{deluxetable*}

\begin{table}
\renewcommand{\thetable}{\arabic{table}}
\centering
\caption{Grid of parameters used for the total SEDs of type-2 AGNs with 
{\tt CIGALE}.}

\label{T3}
\begin{tabular}{lc}
\tablewidth{0pt}
\hline
\hline
Parameter & Value\\
\hline
\multicolumn2c{Delayed SFH}\\
\hline
$\tau_{\rm main}$ [Myr] & 40, 100, 251, 631 \\
& 1584, 3981, 10000, 25118 \\
age [Myr] & 40, 100, 251, 631 \\
& 1584, 3981, 10000 \\
\hline
\hline
\multicolumn2c{Dust attenuation modified starburst}\\
\hline
$E(B-V)_{\rm lines}$ & 0.01, 0.2, 0.5, 0.8, 1.0 \\
Extinction law & 1, 3 \\
\hline

\multicolumn2c{Dust emission \citep{Dale}}\\
\hline
IR power-law slope ($\alpha_{\rm dust}$) & 0.0625, 1.5000, 2.0000, 4.0000 \\
\hline
\multicolumn2c{AGN emission \citep{Stalevski16}}\\
\hline
$\tau_{\rm 9.7}$ 			& 	3, 5, 9		\\
$p$							&	0.0, 1.0, 1.5	\\
$q$							&	0.0, 1.0, 1.5	\\
$\Delta$					&	10, 30, 50\arcdeg	\\
$R_{\rm max}/R_{\rm min}$ 	& 	30 			\\
$\theta$					&	70\arcdeg	\\
$f_{\rm AGN}$ 				& 	0.1, 0.3, 0.5, 0.7, 0.9, 0.99\\
\hline
\multicolumn2c{AGN polar dust emission \citep{yang2020x}}\\
\hline
$E(B-V)$                    & 	0.0, 0.2, 0.4	\\
$T_{\rm dust}^{\rm ploar}$	&	100.0, 1000.0	\\
\hline
\multicolumn2c{Xray emission}\\
\hline
$\alpha_{\rm ox}$			&	-1.9, -1.8, -1.7, -1.6, -1.5\\
& -1.4, -1.3, -1.2, -1.1	\\
max dev $\alpha_{\rm ox}$	&	0.0 \\
\hline
\end{tabular}
\end{table}

\begin{figure*}[ht!]
 \gridline{\fig{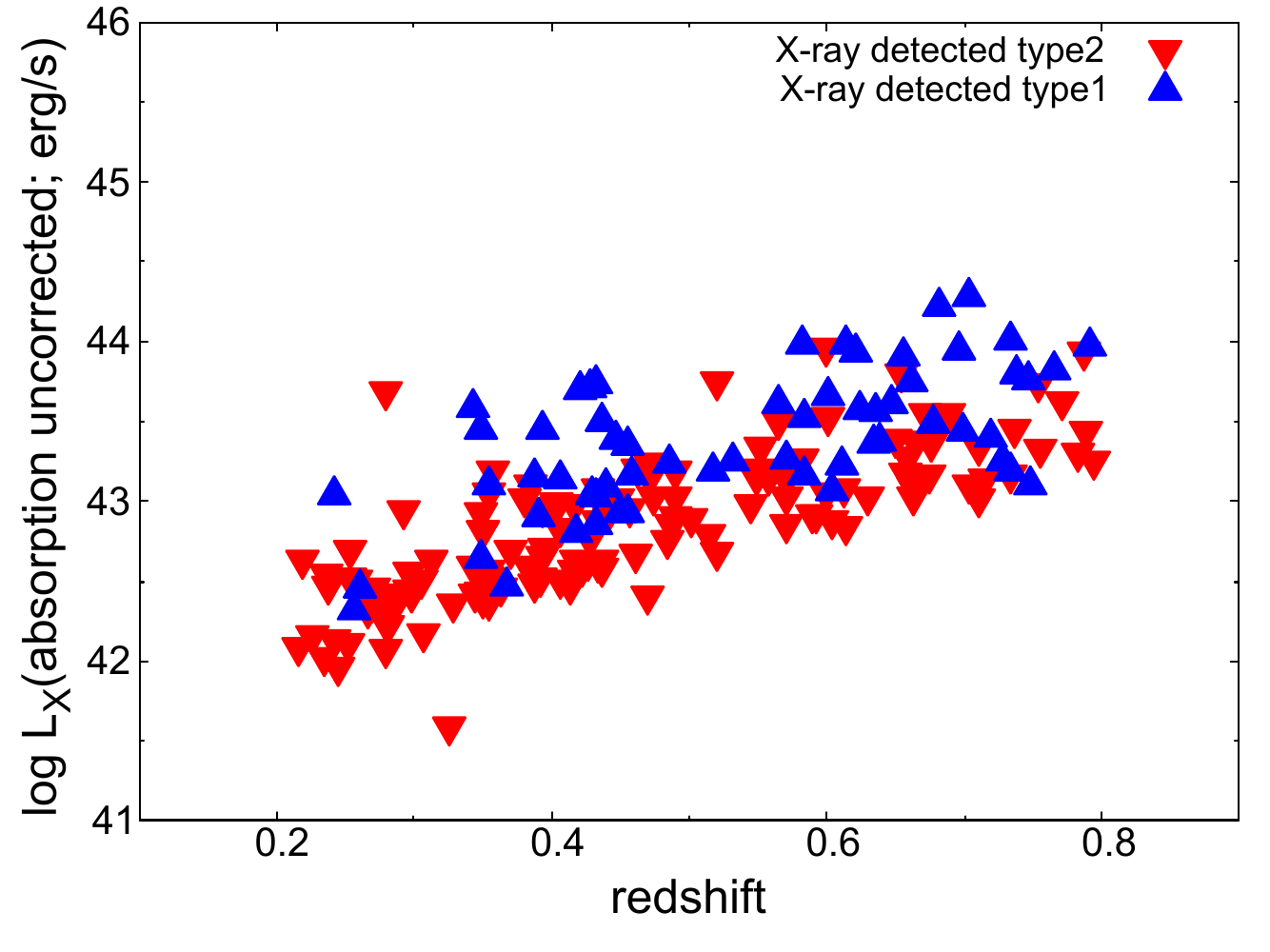}{0.5\textwidth}{(a)}\fig{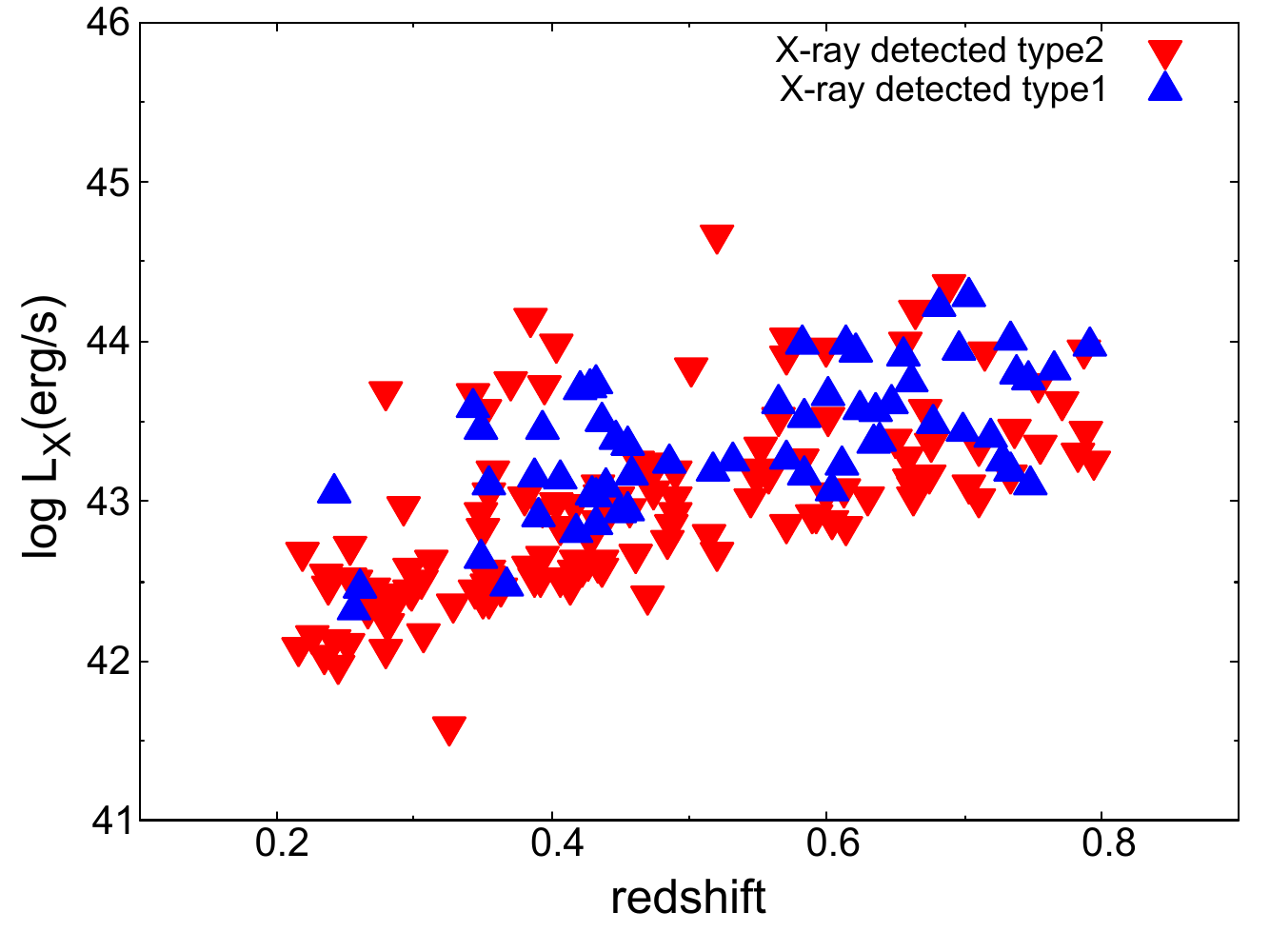}{0.5\textwidth}{(b)}
 }
 \caption{Plot of (a) observed (absorption uncorrected) and (b) intrinsic (absorbed corrected)} 2--10 keV luminosities ($L_{\rm X}$) versus redshift for our type-1 AGNs (blue triangles) and type-2 AGNs (red inverse triangles).
 \label{fig:z_lx}
\end{figure*}

\startlongtable
\begin{deluxetable*}{ccccc}
\tablecaption{Summary of properties of X-ray detected type-2 AGNs and hosts}
\tablewidth{0pt}
\tablehead{
\colhead{ID (Stripe 82X)} & \colhead{SpecObjID} & \colhead{redshift} & \colhead{log $L_{\rm bol}\ ({\rm erg\ s}^{-1})$} & \colhead{log $M_{\rm stellar} \ (M_\odot)$} }
\decimalcolnumbers
\startdata 
\hline
304 & 1683335858842789888 & 0.257 & 44.09 & 10.61 $\pm$ 0.02\\
439 & 9896847496532959232 & 0.325 & 44.21 & 10.52 $\pm$ 0.04\\
2420 & 9894536323041042432 & 0.552 & 44.45 & 10.90 $\pm$ 0.05\\
2425 & 9894555014738714624 & 0.286 & 44.16 & 10.52 $\pm$ 0.05\\
2436 & 9894537972308484096 & 0.266 & 43.78 & 10.20 $\pm$ 0.09\\
2500 & 9894515157442207744 & 0.226 & 43.31 & 10.95 $\pm$ 0.06\\
2536 & 9894517631343370240 & 0.417 & 44.30 & 10.94 $\pm$ 0.10\\
2563 & 9894504986959650816 & 0.305 & 44.08 & 9.60 $\pm$ 0.22\\
2596 & 9894502787936395264 & 0.755 & 44.91 & 10.94 $\pm$ 0.10\\
2605 & 9894598720325918720 & 0.589 & 44.12 & 10.07 $\pm$ 0.10\\
2609 & 9894493442087559168 & 0.462 & 44.48 & 10.77 $\pm$ 0.06\\
2634 & 8865474801944207360 & 0.754 & 45.75 & 10.91 $\pm$ 0.13\\
2644 & 9894489868674768896 & 0.471 & 44.24 & 10.61 $\pm$ 0.10\\
2702 & 9894620160802660352 & 0.484 & 44.38 & 10.73 $\pm$ 0.06\\
2716 & 9894622359825915904 & 0.349 & 45.37 & 9.72 $\pm$ 0.17\\
\hline 
\enddata
\tablecomments{(1) Unique identifier in the Stripe 82X catalog. (2) Unique spectrum identifier in SDSS DR14. (3) Redshift from the SDSS spectra. (4) Logarithmic bolometric AGN luminosity derived by CIGALE, for which we adopt the output parameter ``agn.accretion
power''. (5) Logarithmic stellar mass derived by CIGALE and its 1$\sigma$ error.
(This table is available in its entirety in machine-readable form.)}
\label{table:parameter_catalog}
\end{deluxetable*}

\subsection{X-ray Luminosities}
\label{subsec:Lx}

Figure~\ref{fig:z_lx} (a) plots
observed (absorption uncorrected) rest-frame 2--10 keV luminosity versus redshift 
for our type-1 and type-2 AGN samples,
consisting of 60 and 137 objects, respectively.
We also plot
intrinsic (absorption corrected) luminosity ($L_{\rm X}$) versus redshift for the same samples
in Figure~\ref{fig:z_lx} (b).
We calculated $L_{\rm X}$ by correcting for absorption if
present, using the hardness ratio between the 0.5-2 keV and 2--10 keV
(for XMM-Newton) or 2--7 keV (for Chandra) fluxes available in the
Stripe-82X catalog. Following the recipe in \cite{Ueda03}, we assume a
power law photon index of 1.9 and take into account a reflection
component from cold matter covering a solid angle of $2\pi$. As
noticed from the figure, the luminosity range spans $41.6<\log L_{\rm X}<44.7$, covering the low luminosity range (log $L_{\rm X}<43.5$) we are
particularly interested in (see Section \ref{sec:intro}).
The mean and standard deviation of log $L_{\rm
  X}$ are $43.38\pm0.43$ for the type-1 AGNs and
$42.94\pm0.56$ for the type-2 AGNs.

Here we compare our samples with those previously studied in
the Stripe 82 field. \cite{LaMassa16b} investigated the optical to
mid-infrared colors of 552 X-ray selected AGNs
with WISE and UKIDSS detections, based on the previous version of Stripe 82X catalog
utilizing the XMM-Newton AO10 data. Their sample contains 24 type-1 and 3 type-2 AGNs in our
sample. \cite{LaMassa19} listed 4847 AGN candidates based on the X-ray
and WISE data. The overlap with our sample is 36 out of the 60 type-1
AGNs and 25 out of the 137 type-2 AGNs. We have confirmed that the
luminosity ranges of the overlapping objects are similar to those of
our samples. \cite{LaMassa17} and \cite{Glikman18} studied the properties of
12 ``red quasar'' candidates and 147 WISE selected AGNs, respectively.
None of these are included in our samples. This point will be discussed in
Section \ref{subsec:z_Lbol_massratio}.

\subsection{New Black Hole Mass Estimation in Type-1 AGNs Based on Image Decomposition}
\begin{figure*}[ht!]
\gridline{\fig{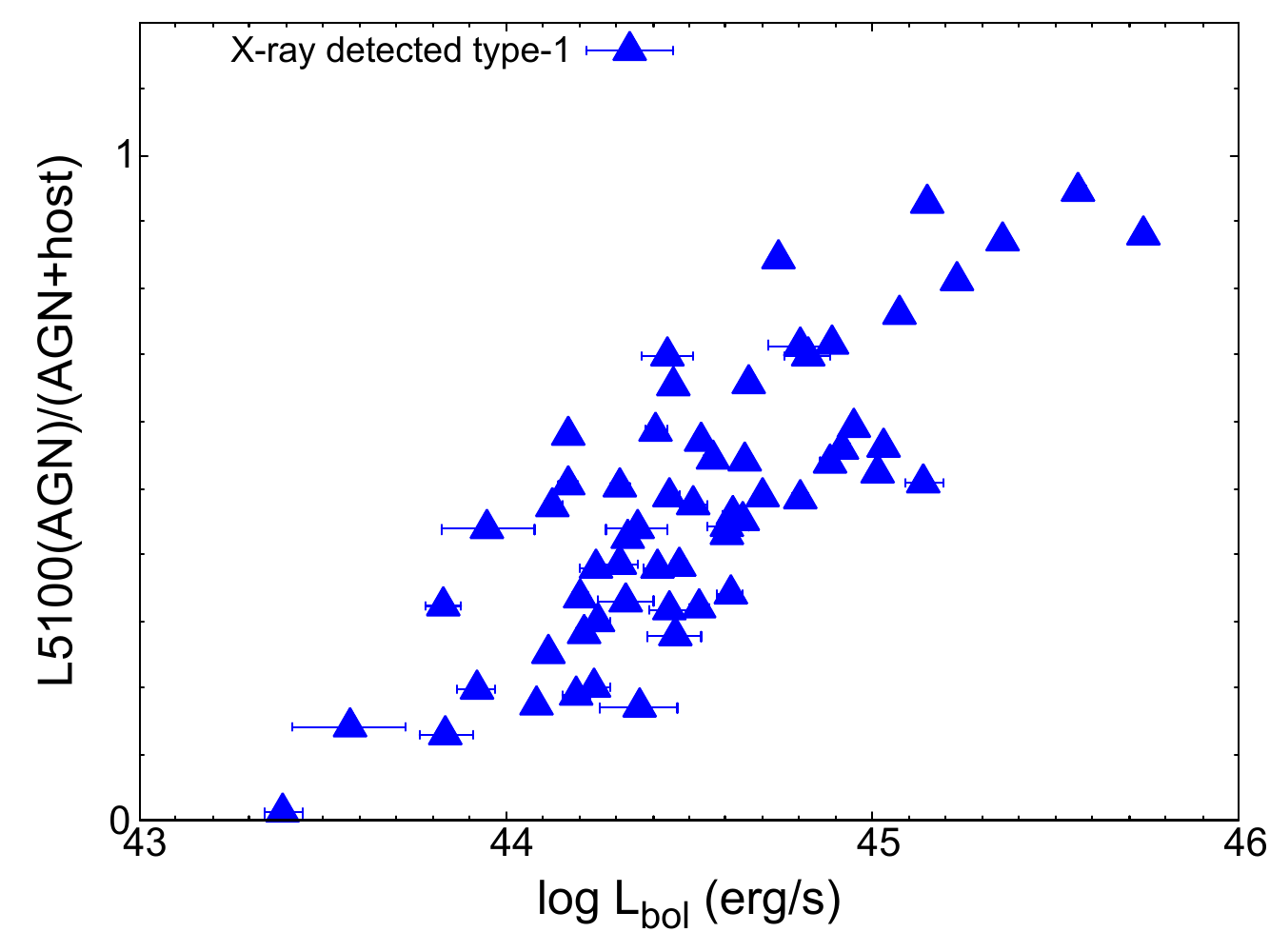}{0.5\textwidth}{(a)}\fig{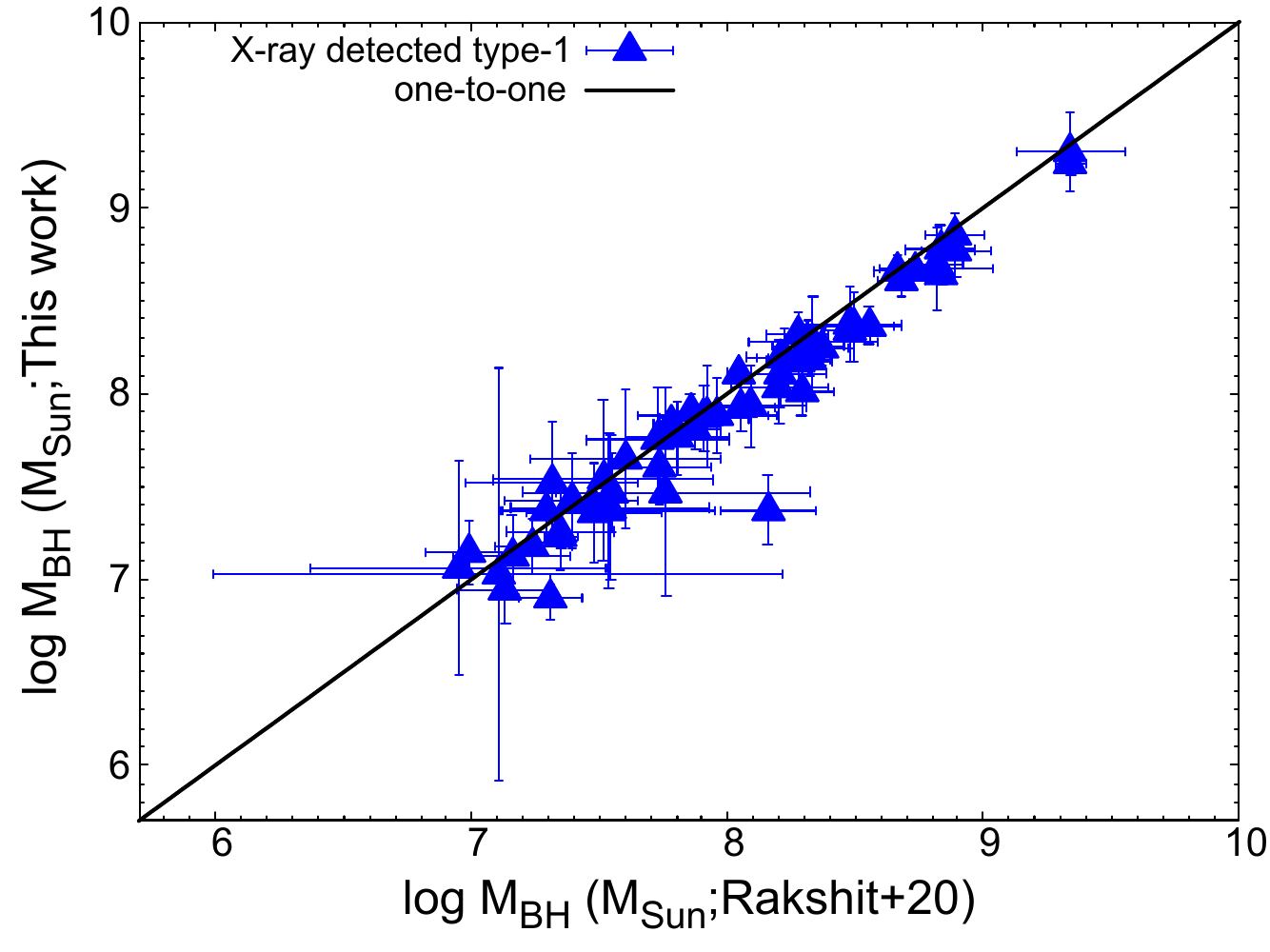}{0.5\textwidth}{(b)}}
\caption{(a) Fractional AGN contribution to the total continuum luminosity at 5100\AA\ as a function of $L_{\rm bol}$. (b) The comparison of $M_{\rm BH}$ between our new estimates and those by \cite{Rakshit20}. The black solid line shows the one-to-one relation. The blue triangles represent X-ray detected type-1 AGNs.}
\label{fig:L5100_MBHestimate}
\end{figure*}

To reliably estimate the SMBH mass of a type-1 AGN by using the broad
line width and continuum luminosity ($L_{\rm 5100}$), it is critical to properly subtract the host contribution at rest-frame 5100\AA,
particularly at a low AGN luminosity. To decompose the host galaxy
contribution in the optical spectrum, \cite{Rakshit20} performed the
PCA using solely spectral information. In our work, we are able to accurately
separate the AGN and host components at 5100\AA\ on the basis of
image decomposition and multiwavelength SED fitting. Figure~\ref{fig:L5100_MBHestimate}(a)
plots the fraction of AGN contribution to the total rest-frame 5100\AA\ luminosity as a function of AGN bolometric luminosity, showing that
the host contamination becomes more significant toward lower AGN
luminosities. We compare the SMBH masses obtained with our method and
those by \cite{Rakshit20} in Figure~\ref{fig:L5100_MBHestimate}(b). It is seen that we obtain
slightly smaller SMBH masses than those in \cite{Rakshit20}.
Throughout this work, we adopt the SMBH masses estimated by our method, which are also listed in Table \ref{table:parameter_catalog}.
The Eddington ratio is calculated as $\lambda_{\rm Edd} = L_{\rm
  bol}$/$L_{\rm Edd}$, where $L_{\rm Edd} = 1.3\times10^{38} M_{\rm
  BH}/M_\odot$.

\section{Results and Discussion} \label{sec:resultsanddiscussion}


\subsection{type-1 AGNs}

\subsubsection{Statistics of AGN and Host Parameters}

\begin{figure*}[ht!]
\gridline{\fig{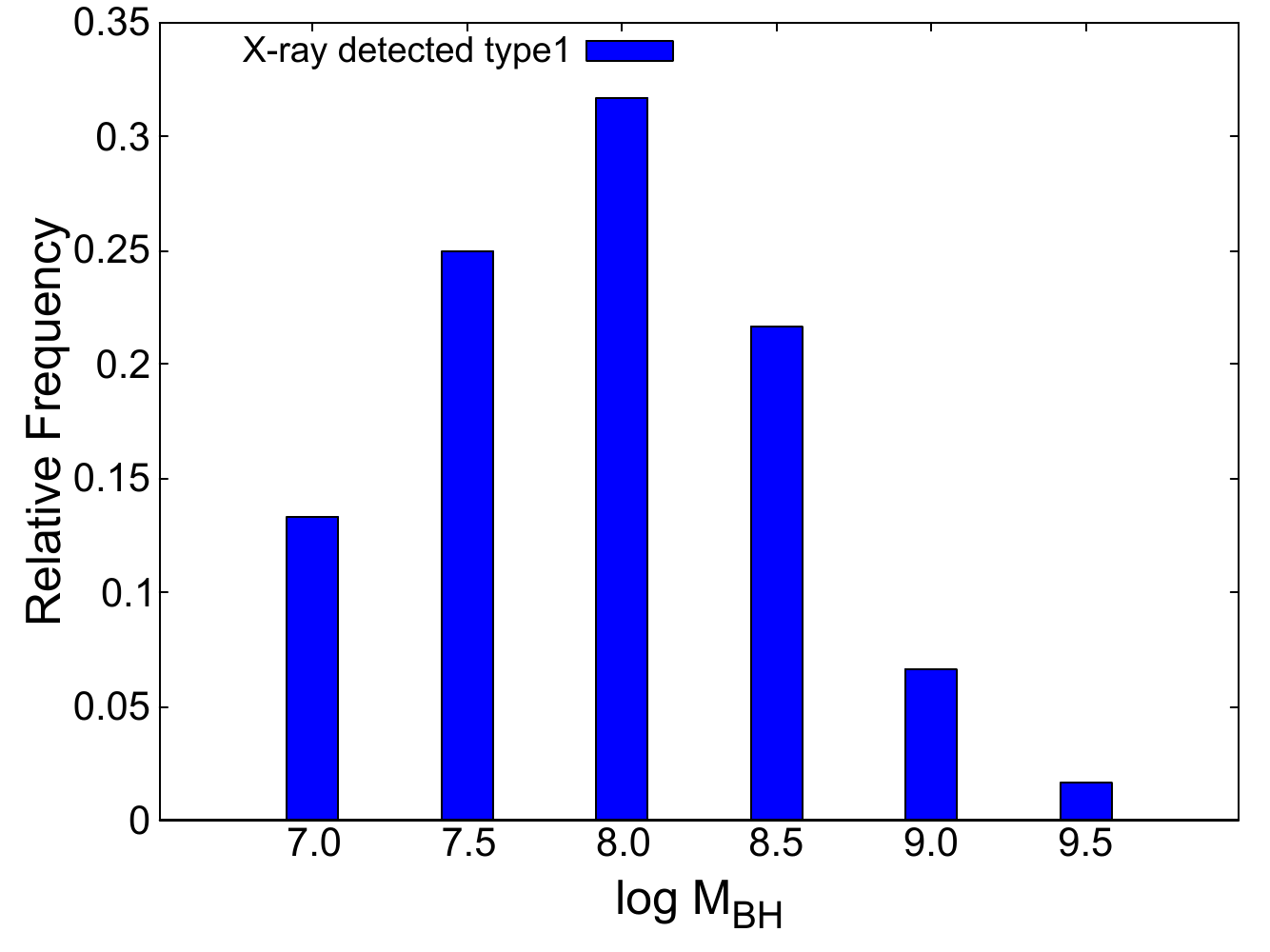}{0.35\textwidth}{(a)}\fig{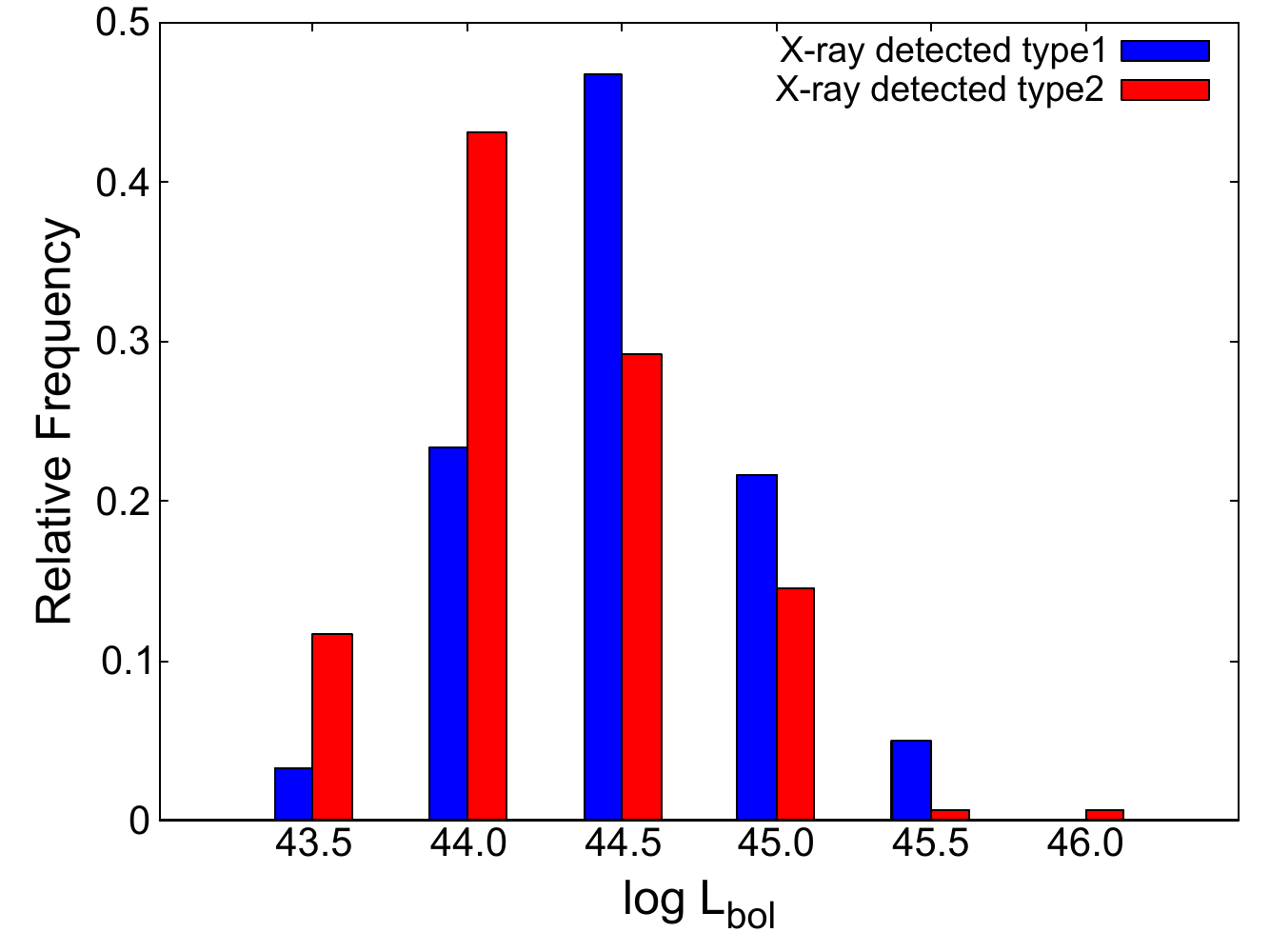}{0.35\textwidth}{(b)}
\fig{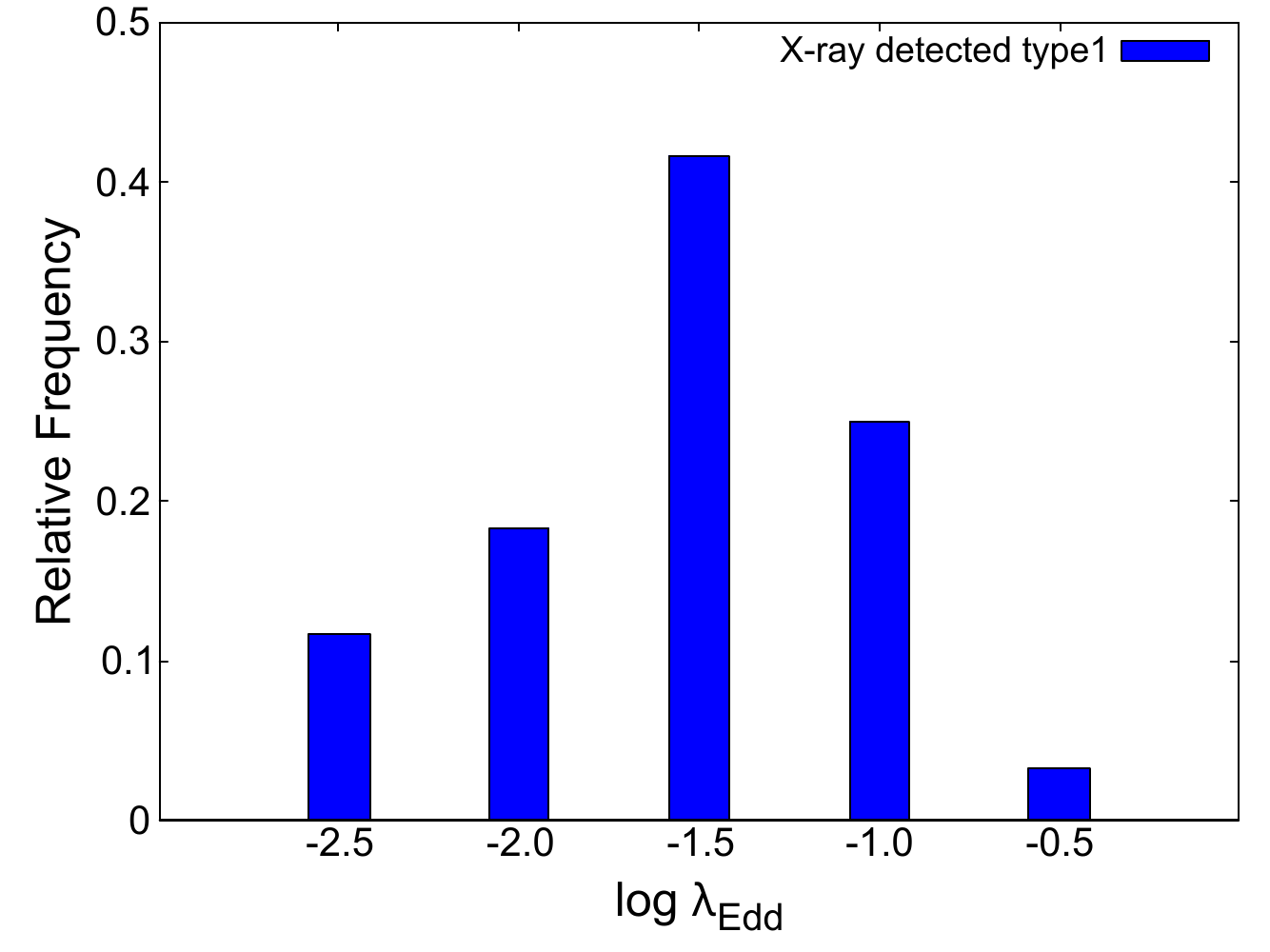}{0.35\textwidth}{(c)}}
\gridline{\fig{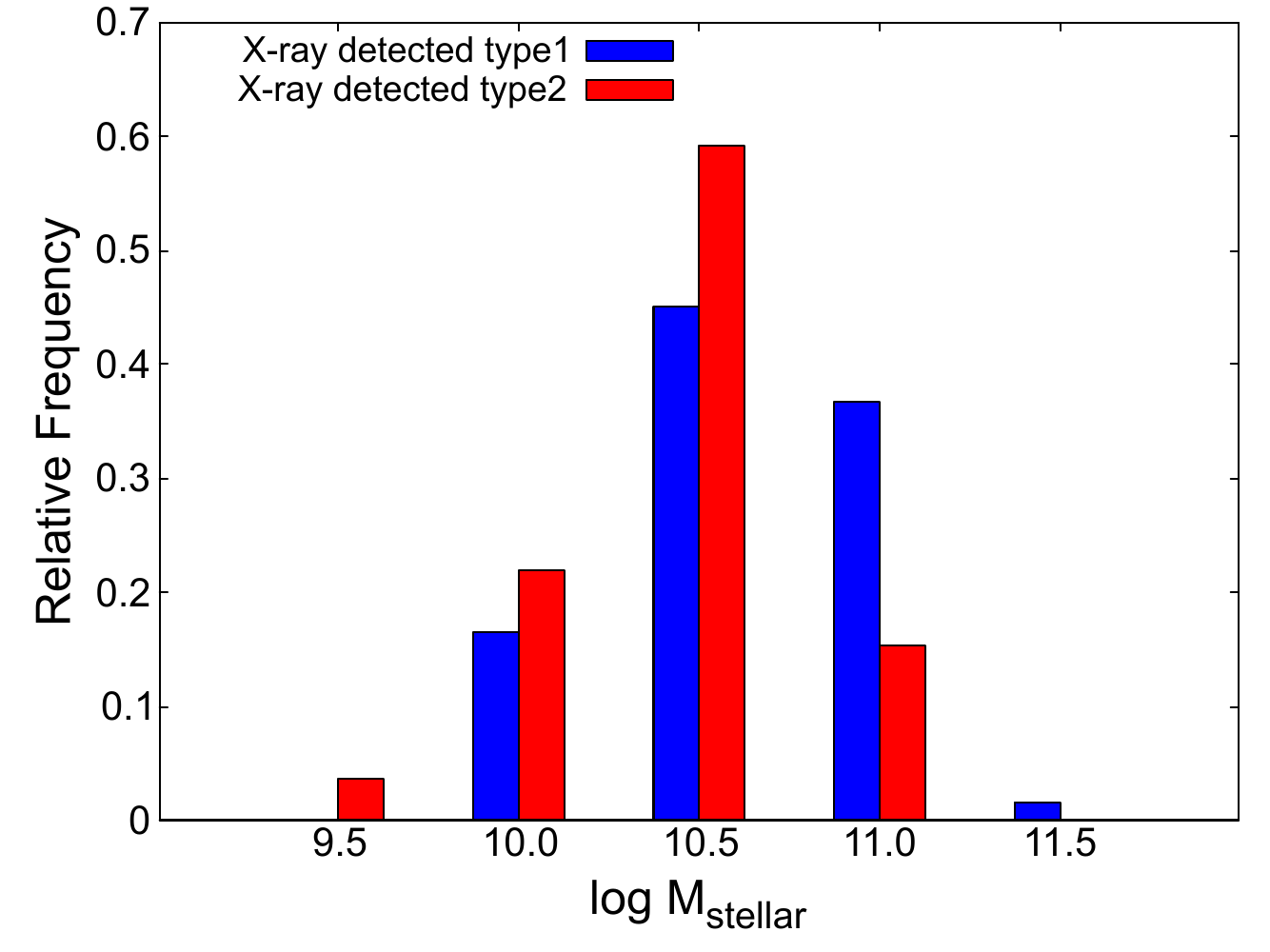}{0.35\textwidth}{(d)}
          \fig{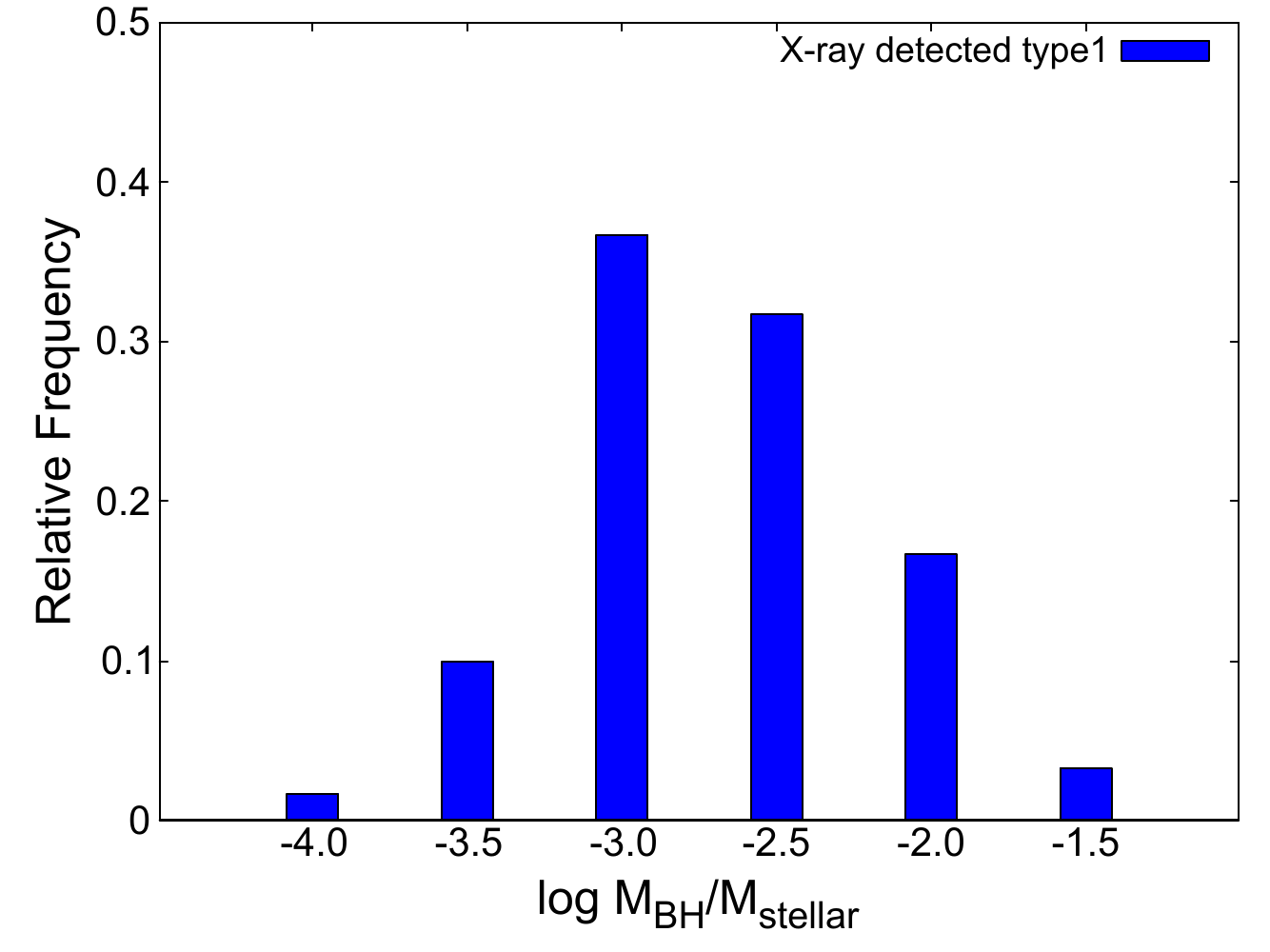}{0.35\textwidth}{(e)}
          \fig{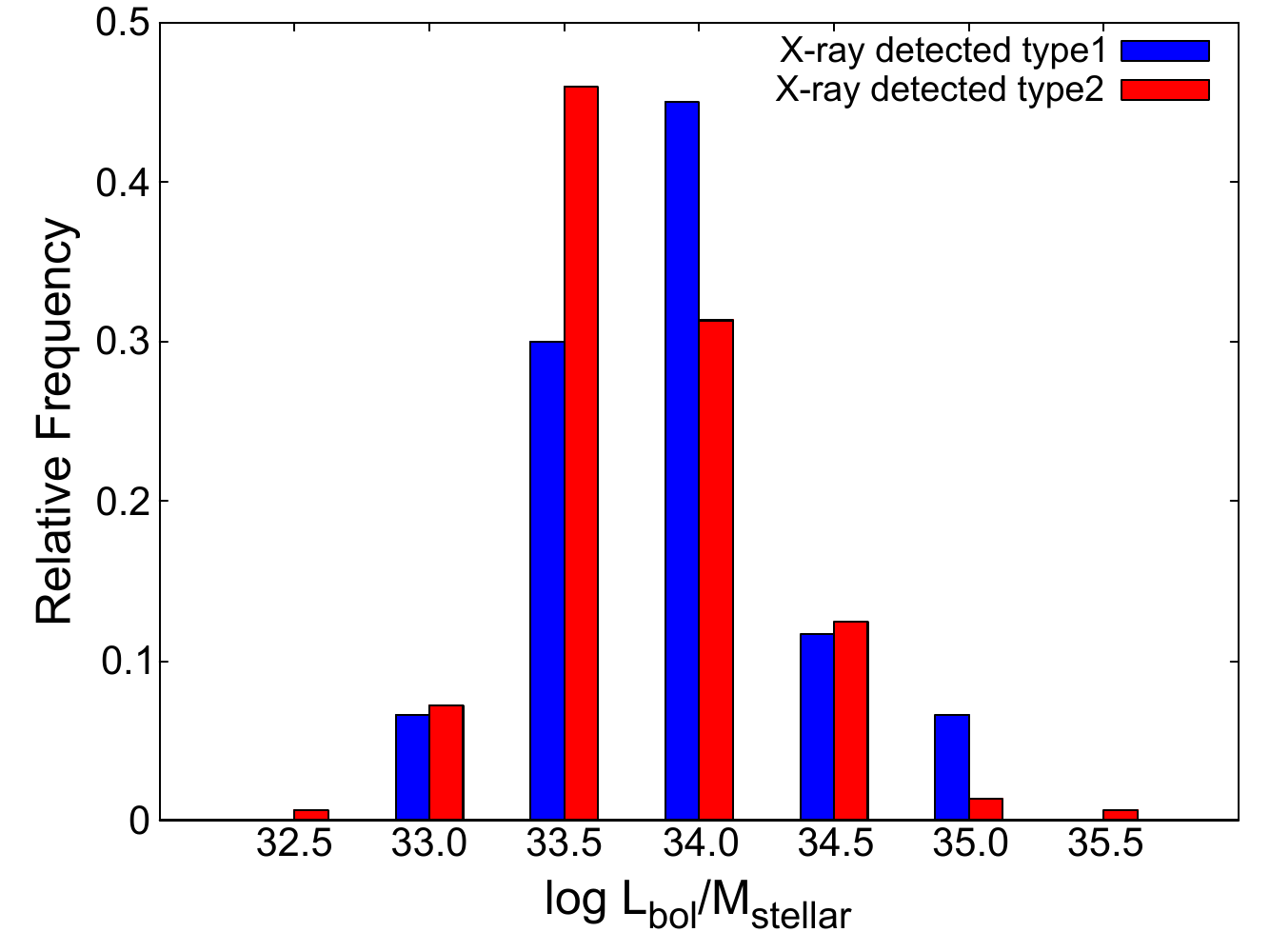}{0.35\textwidth}{(f)}}
\caption{Histograms of (a) log $M_{\rm BH}$ [$M_\odot$], (b) log $L_{\rm bol}$ [erg $s^{-1}$], (c) log $\lambda_{\rm Edd}$, (d) log $M_{\rm stellar}$ [$M_\odot$], (e) log $(M_{\rm BH}/M_{\rm stellar})$, and (f) log specific Eddington ratio ($L_{\rm bol}/M_{\rm stellar}$). The blue and red histograms correspond to type-1 and type-2 AGNs, respectively.}
\label{fig:histogram}
\end{figure*}

Figure \ref{fig:histogram} plots the histograms of the best-fit parameters of the {\tt
  CIGALE} SED fitting for our type-1 AGN sample:
(a) $M_{\rm BH}$, (b) $L_{\rm bol}$, (c) $\lambda_{\rm Edd}$, (d) $M_{\rm stellar}$, (e) $M_{\rm BH}/M_{\rm stellar}$, and (f)
$L_{\rm bol}/M_{\rm stellar}$. The median, mean, and standard deviation of these parameters in type-1 AGNs are listed in
Table~\ref{table:output_parameter}. Our sample covers 68\% ($\pm 1\sigma$) regions
of log $M_{\rm BH}/M_\odot = 7.83\pm0.64$, log $L_{\rm bol}/ {\rm erg\ s}^{-1} = 44.52\pm0.45$, and log $M_{\rm stellar}/ M_\odot = 10.61\pm0.35$.
This is one of the largest X-ray selected type-1 AGN samples with
reliable stellar mass estimates covering a low to medium luminosity
range of log $L_{\rm bol}<45$ at $z=0.2-0.8$.
We discuss the differences between the type-1 and type-2 samples in Section \ref{sec:control}.
\begin{deluxetable}{cccc}
  \tablecaption{Statistical properties of key parameters in our type-1 and type-2 AGN samples.}
\tablewidth{0pt}
\tablehead{
\colhead{Parameter} & \colhead{Median} & \colhead{Mean} & \colhead{$\sigma$}}
\decimalcolnumbers
\startdata 
\multicolumn2c{X-ray detected type-1 (main sample)}\\
\hline
log $M_{\rm BH}$ [$M_{\odot}$]& 7.91 & 7.93 & 0.58\\
log $M_{\rm stellar}$ [$M_{\odot}$]& 10.68 & 10.61 & 0.35 \\
log $L_{\rm bol}$ [${\rm erg}\ s^{-1}$]& 44.47 & 44.52 & 0.45 \\
log $\lambda_{\rm Edd}$ & -1.49 & -1.52 & 0.47 \\
log $(M_{\rm BH}/M_{\rm stellar})$ & -2.67 & -2.68 & 0.50 \\
log $(L_{\rm bol}/M_{\rm stellar})$ [${\rm erg}\ s^{-1}\ M_{\odot}^{-1}$] & 33.88 & 33.91 & 0.44 \\
\hline
\multicolumn2c{X-ray detected type-2}\\
\hline
log $M_{\rm stellar}$ & 10.52 & 10.46 & 0.34 \\
log $L_{\rm bol}$ & 44.20 & 44.25 & 0.46 \\
log $(L_{\rm bol}/M_{\rm stellar})$ [${\rm erg}\ s^{-1}\ M_{\odot}^{-1}$] & 33.70 & 33.80 & 0.44 \\
\hline
\enddata
\tablecomments{(1) Key AGN and host parameters in our type-1 and type-2 AGN samples. Statistical properties are summarized in terms of (2) median, (3) mean, and (4) $\sigma$ (standard deviation).}
\label{table:output_parameter}
\end{deluxetable}


\subsubsection{Relations among $M_{\rm BH}$, $L_{\rm bol}$, $M_{\rm stellar}$}
\label{subsubsec:MBH_Lbol_Mstellar}

\begin{figure*}[ht!]
\gridline{\fig{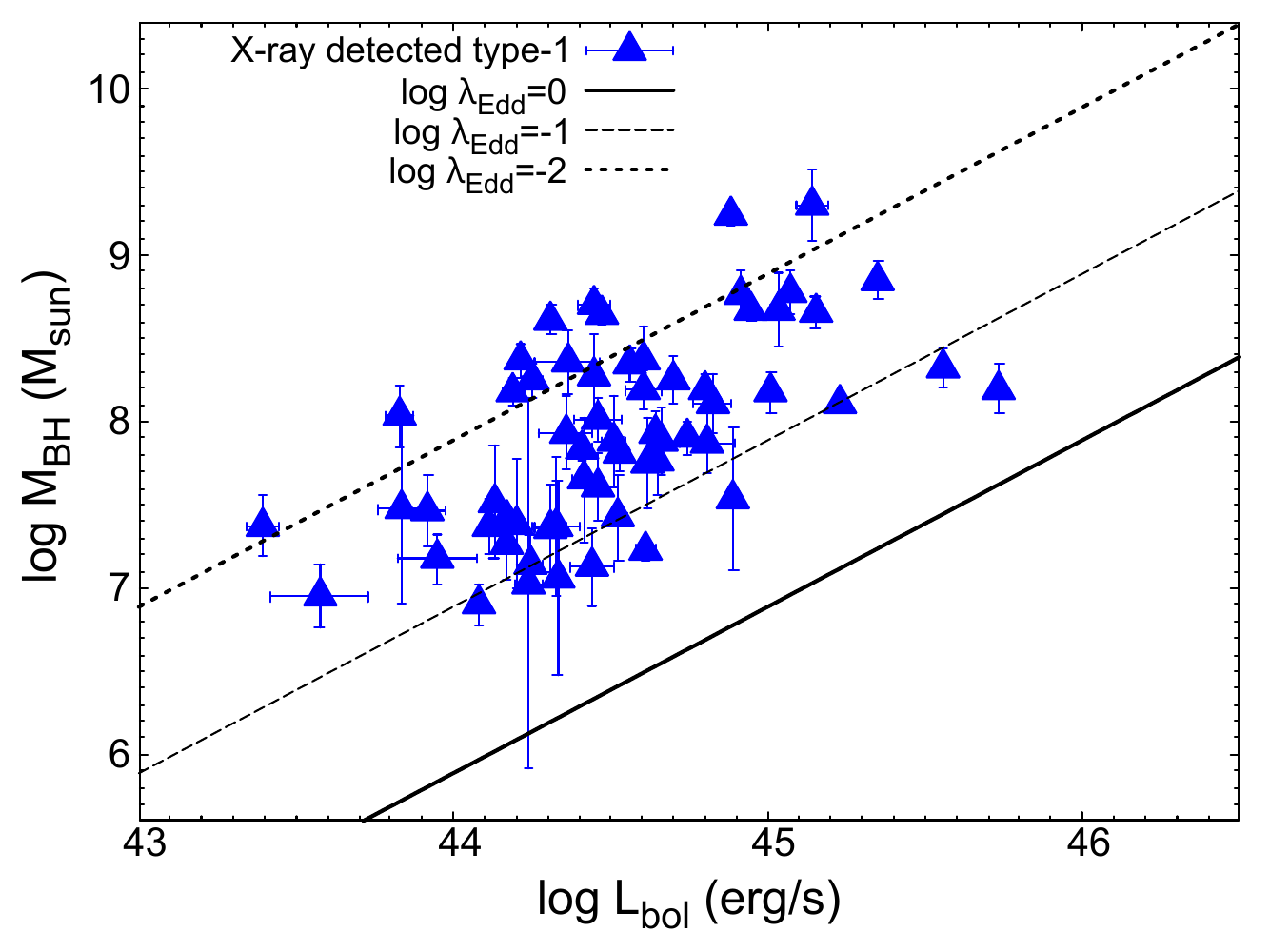}{0.45\textwidth}{(a)}\fig{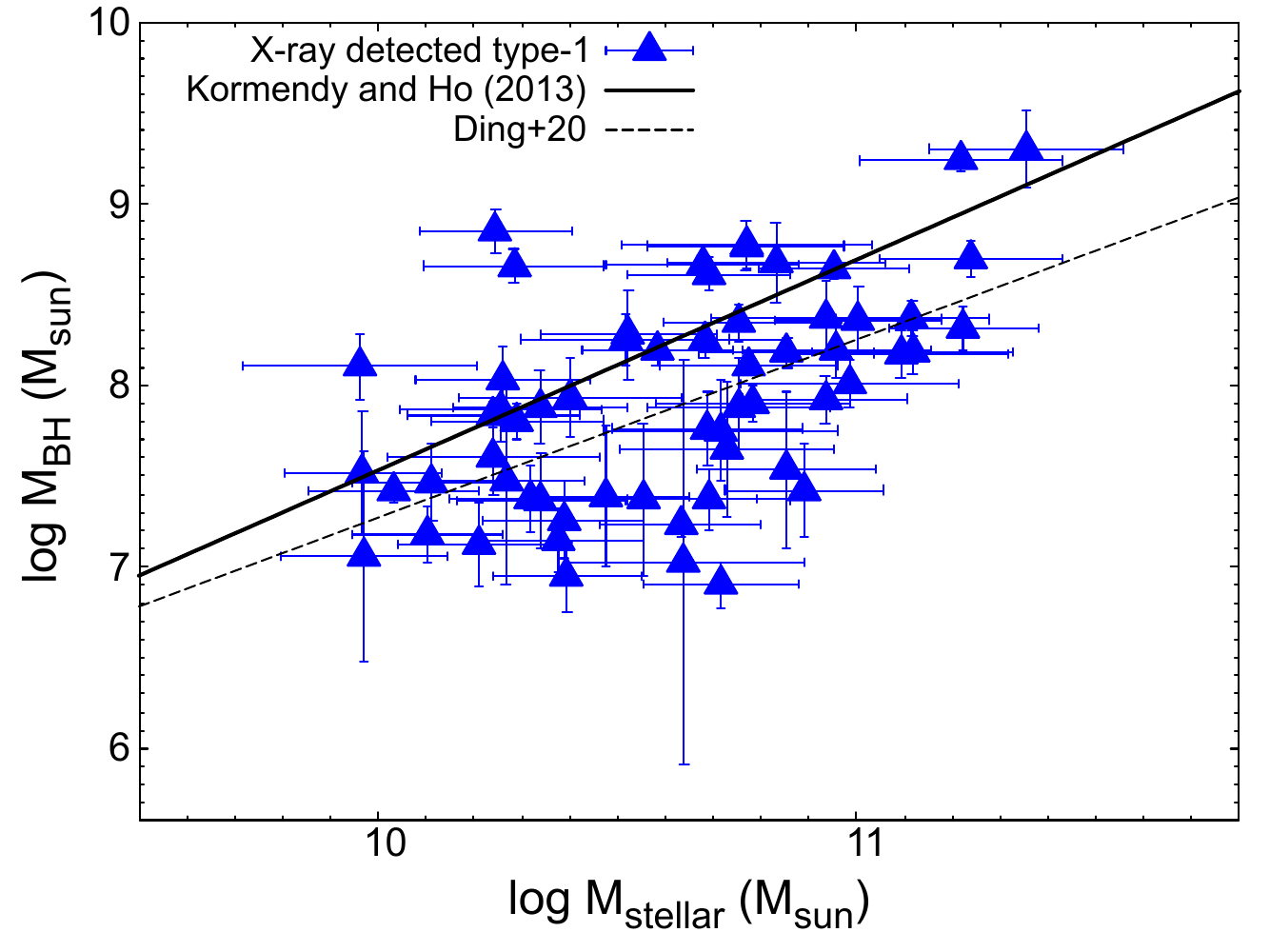}{0.45\textwidth}{(b)}}
\gridline{\fig{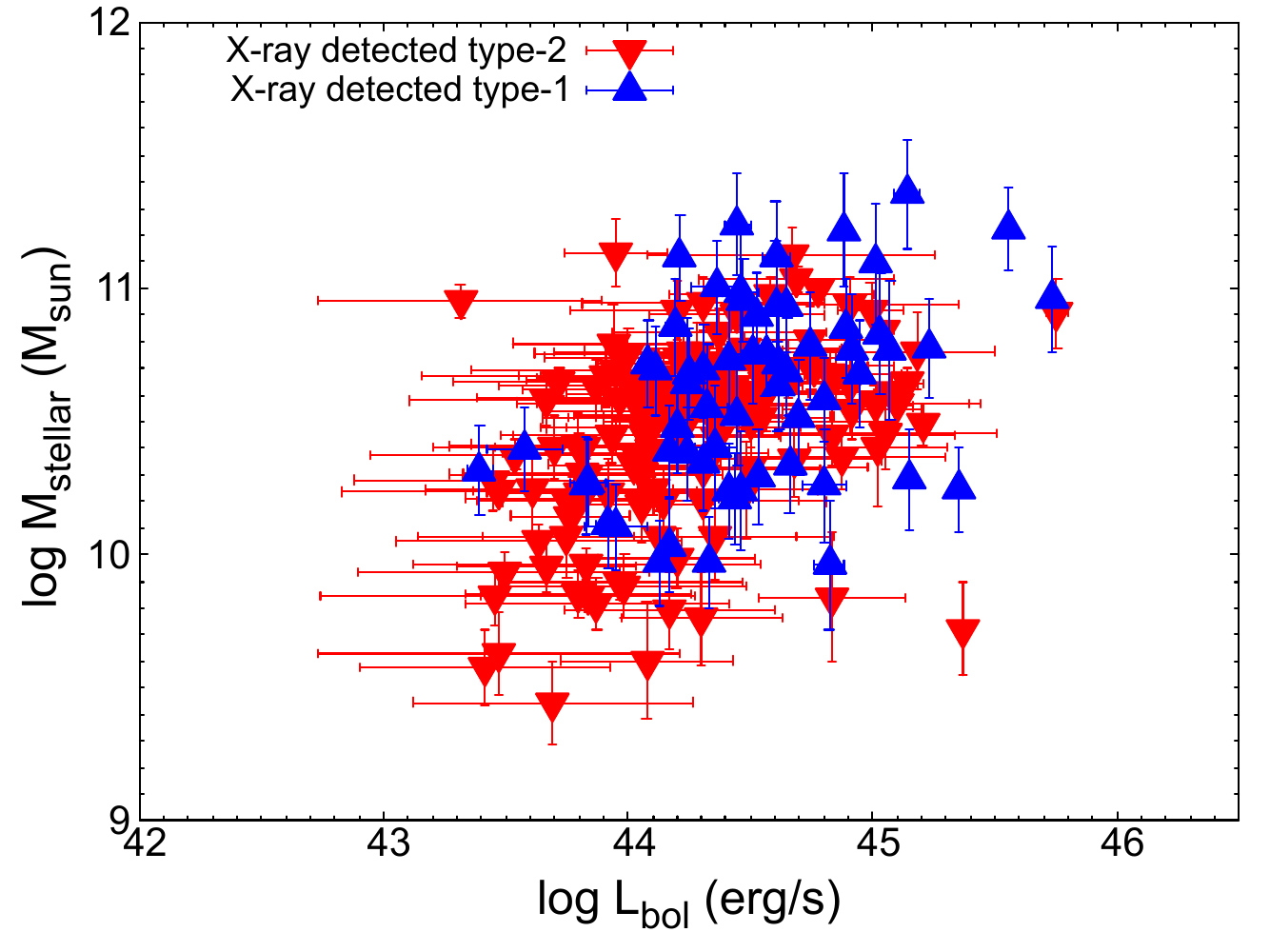}{0.45\textwidth}{(c)}}
\caption{Relations between (a) $M_{\rm{stellar}}$ and
    $M_{\rm{BH}}$, (b) $L_{\rm{bol}}$ and
    $M_{\rm{BH}}$, and (c)$M_{\rm{stellar}}$ and
  $L_{\rm{bol}}$.
  (a) The black solid and dashed lines show the Eddington ratio $(\lambda_{\rm Edd})=0, -1, \rm{and}-2$. (b) The black solid and dashed 
     lines represent the local black hole-to-bulge mass relation from
     \cite{Kormendy} and from \cite{Ding20}, respectively.
}
\label{fig:coevolution_parameters}
\end{figure*}

Figure~\ref{fig:coevolution_parameters} (a) plots the relation between
$M_{\rm BH}$ and $L_{\rm bol}$ of our sample.
We plot constant Eddington-ratio lines corresponding to log
$\lambda_{\rm Edd}$ = --2.0, --1.0, and 0.0. As noticed, most of our
objects are distributed between
log $\lambda_{\rm Edd}$= --2.5 and --0.5 with a mean value of --1.5 (Table \ref{table:output_parameter}).
Since the scatter in $M_{\rm BH}$ is larger than in $\lambda_{\rm
  Edd}$ (Figure \ref{fig:histogram} (a) and (c)), we may regard that the bolometric luminosity, which is the product of $M_{\rm BH}$ and $\lambda_{\rm Edd}$, is mainly determined by $M_{\rm BH}$ in our sample.
It is beyond the scope of this paper to derive the intrinsic Eddington
ratio distribution function (ERDF) by correcting for all sample
selection biases. Nevertheless, the peak in the observed
distribution, $\log \lambda_{\rm Edd} \sim -1.5$, is similar to that
found in the local hard X-ray selected type-1 AGN sample \citep{Koss17},
implying that the ERDF of type-1 AGNs little evolves from $z<0.2$ to
$z=0.2-0.8$.

Figure \ref{fig:coevolution_parameters} (b) plots the relation between
$M_{\rm{stellar}}$ and $M_{\rm{BH}}$. We perform a correlation analysis
with the method of \cite{Kelly07}, which allows us to take into account
the parameter errors. We obtain a correlation
coefficient of $r=0.64\pm0.12$, indicating a positive correlation.
This supports that $M_{\rm{stellar}}$ may be used as a proxy of
$M_{\rm{BH}}$, at least for type-1 AGNs, albeit with a $1\sigma$ scatter of 0.5 dex (Table~\ref{table:output_parameter}).
The relation between $L_{\rm{bol}}$ and $M_{\rm{stellar}}$ is shown
in Figure \ref{fig:coevolution_parameters}~(c). A similar positive
correlation as found between $L_{\rm{bol}}$ and $M_{\rm{BH}}$ in
Figure \ref{fig:coevolution_parameters}~(a) is noticed. This is
expected because $M_{\rm{stellar}}$ and $M_{\rm{BH}}$ are correlated.
In Figure~\ref{fig:coevolution_parameters}(b), we display the 
local $M_{\rm{BH}}$-$M_{\rm{bulge}}$ relation obtained by \cite{Kormendy} 
and by \cite{Ding20}.
The mean value of log $(M_{\rm{BH}}/M_{\rm{stellar}})$ is found to be
$-2.7\pm0.5$ (Table~\ref{table:output_parameter}). This result is 
consistent with the earlier report by \cite{Li21b} for SDSS type-1 quasars.
This is expected because our X-ray selected type-1 AGN sample is a
sub-sample of \cite{Li21b}'s one and X-ray detection causes no
significant selection biases between them (see Appendix~A).
\ifnum0=1
Since \cite{Li21b} finds no significant evolution of $(M_{\rm BH}/M_{\rm stellar})$ with
redshift between $z=0.2$ and $z=0.8$, hereafter we ignore any redshift dependence at $z=0.2-0.8$.
\fi
This mean value of $M_{\rm{BH}}/M_{\rm{stellar}}$ in our sample is similar to the local SMBH-to-bulge mass ratio, log
$(M_{\rm{BH}}/M_{\rm{bulge}})=-2.4$.  We find that all of our host
galaxies have S\'{e}rsic index $n<2.5$ in the \cite{Li21a} catalog,
and hence are likely to have disk-dominant morphologies.  Since
$M_{\rm{stellar}}$ includes $M_{\rm{bulge}}$ and galactic disk mass,
it is suggested that the $M_{\rm{BH}}-M_{\rm{bulge}}$ ratio in our
sample should be larger than the local value, that is, our objects
have overmassive black holes relative to galactic bulges. As discussed
in e.g., \cite{Dekel14,Shangguan20,Li21a}, in a later stage of the AGN
phase currently observed in our sample, the concentrated gas through
reservoirs or gas compaction mechanisms (e.g., minor mergers or disk
instabilities) must take place to enhance star formation in classical
bulges to overtake the $M_{\rm BH}$ evolution.

\subsubsection{Redshift and Luminosity Dependence of $(M_{\rm BH}/M_{\rm stellar})$}
\label{subsec:z_Lbol_massratio}

\begin{deluxetable*}{ccccccc}
\tablecaption{Summary of previous studies on the $M_{\rm BH}$-$M_{\rm stellar}$ relation in broad-line AGNs.}
\tablewidth{0pt}
\tablehead{
\colhead{Reference} & \colhead{redshift} & \colhead{log $L_{\rm bol}$} & \colhead{log $(M_{\rm BH}/M_{\rm stellar})$}
& \colhead{morphology} & \colhead{N} & \colhead{X-ray selected}}
\decimalcolnumbers
\startdata 
\hline
This work (log $L_{\rm bol}<45$) & 0.24-0.79 & 44.39-44.95 & -2.75 
& disk-like & 51 & Yes\\
\cite{Schramm13} (all AGN) & 0.544-1.065 & 43.31-45.89 & -2.78
& & 18 & Yes\\
\cite{Schramm13} ($n<2.5$) & 0.57-1.07 & 43.31-45.89 & -2.75 
& disk-like & 12 & Yes\\
\cite{Schramm13} ($n>2.5$) & 0.54-1.04 & 43.31-45.89 & -2.82 
& bulge-like & 6 & Yes\\
\cite{Schramm13} ($B/T<0.5$) & 0.57-1.00 & 43.31-45.89 & -2.71 
& disk-like & 9 & Yes\\
\cite{Schramm13} ($B/T>0.5$) & 0.54-1.07 & 43.31-45.89 & -2.84 
& bulge-like & 9 & Yes\\
\cite{Sun15} (all AGN) & 0.2-2.1 & 43.81-47.15 & -2.64 
& disk-like & 69 & Yes\\
\cite{Sun15} (log $L_{\rm bol}<45$) & 0.2-1.23 & 43.81-44.93 & -3.12 
& disk-like & 16	& Yes\\
\cite{Reines15} & 0-0.06 & 41.5-44.4 & -3.64 
& & 244 & No\\
\cite{Ding20} & 1.24-1.67 & 45.06-46.21 & -2.33 
& & 32 & Yes\\
\cite{Ding20} ($n<2.5$) & 1.24-1.63 & 45.06-46.21 & -2.32 
& disk-like & 19 & Yes\\
\cite{Ding20} ($n>2.5$) & 1.24-1.67 & 45.06-46.21 & -2.33 
& bulge-like & 13 & Yes\\
\cite{Setoguchi21} (all AGN) & 1.18-1.68 & 44.5-46.39 & -2.2 
& & 85 & Yes\\
\cite{Setoguchi21} (log $L_{\rm bol}>45$) & 1.18-1.68 & 45.01-46.39 & -2.1
& & 68 & Yes\\
\hline
\enddata
\tablecomments{(1) References that studied $M_{\rm BH}$-$M_{\rm stellar}$ relation in broad-line AGNs (ordered by publication date). 
$n$ and $B/T$ means the S\'{e}rsic index and the bulge-to-total luminosity ratio, respectively. (2) The redshift range. (3) The log $L_{\rm bol}$ range. (4) The mean value of log $(M_{\rm BH}$/$M_{\rm stellar})$. (5) Morphologies of the host galaxies. (6) The number of objects. (7) If the objects are X-ray selected, this column contains "Yes".}
\label{table:massratio_summary}
\end{deluxetable*}

As we mention in Section \ref{sec:intro}, in order to investigate the origin of
cosmic downsizing of SMBH evolution, it is quite interesting to
compare our results with those obtained for more luminous AGNs at
higher redshifts. Table \ref{table:massratio_summary} summarizes major previous studies
investigating $M_{\rm BH}$-$M_{\rm stellar}$ relations for X-ray selected type-1 AGNs
at various luminosity and redshift ranges. It also provides information on galaxy morphologies whenever available. 
As noticed,
high luminosity AGNs (typically with log $L_{\rm bol}\gtrsim45$) at $z \gtrsim 1.2$ have larger black hole-to-stellar mass ratios, $\log (M_{\rm BH}/M_{\rm stellar}) \sim -2.2$, as compared with lower luminosity ones at
$z \lesssim 1.2$ ($\log (M_{\rm BH}/M_{\rm stellar}) \sim -2.8$ in this work).

\begin{figure}[ht!]
  \plotone{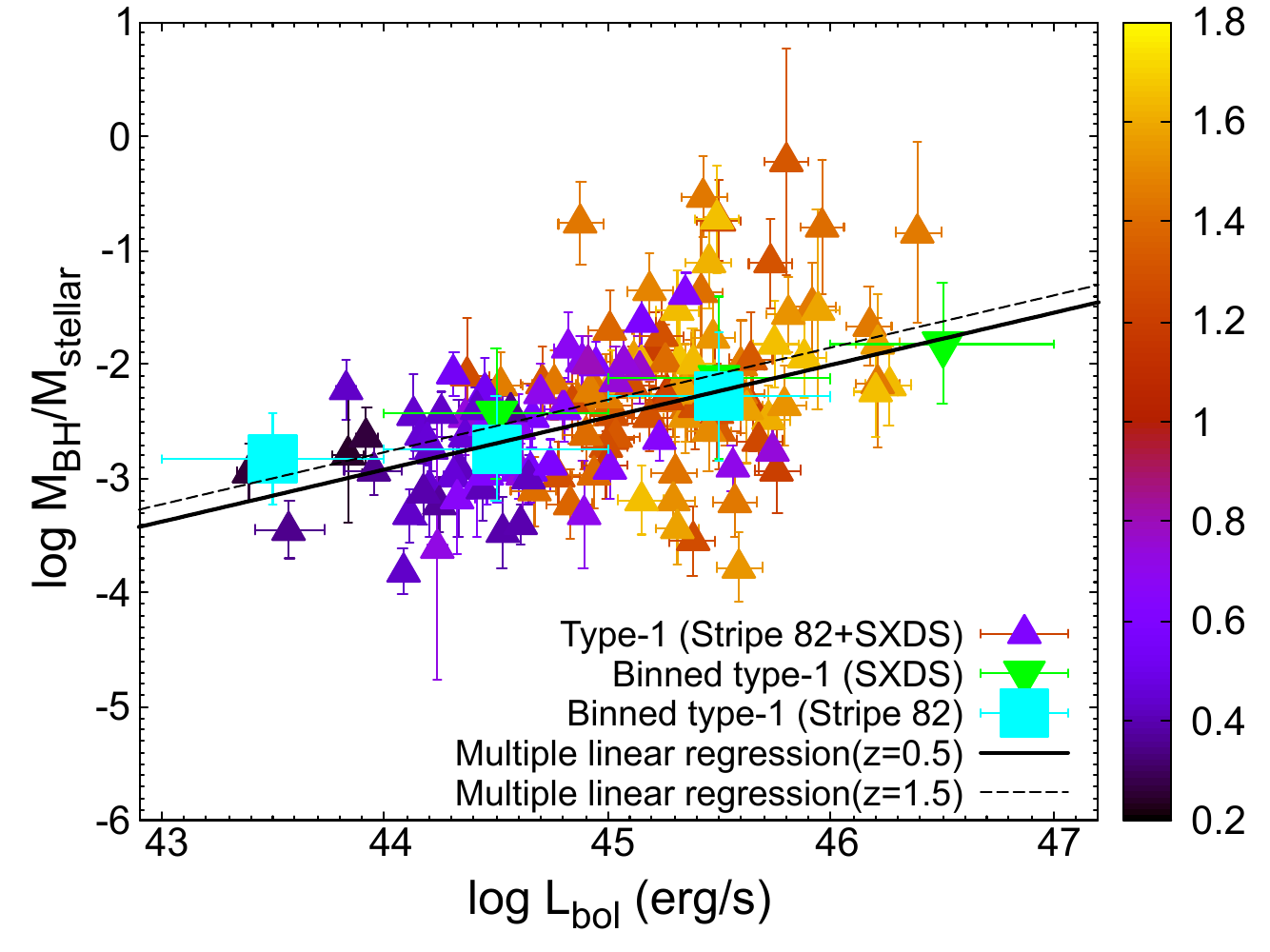}
  \caption{log $(M_{\rm BH}/M_{\rm stellar})$ as a function of log $L_{\rm bol}$.
    The data points are color-coded by redshift. 
    The triangles correspond to X-ray detected type-1 AGNs in the Stripe 82 (this work) and SXDS region \citep{Setoguchi21}. The green inverse triangles and the cyan squares represent bins of objects in the Stripe 82 and SXDS regions, respectively. The vertical position and the error bar show a mean and a standard deviation of log $(M_{\rm BH}/M_{\rm stellar})$ in each bin.
The black solid and dashed lines display the best-fit multiple linear regression in Equation \ref{eq:MLR} at $z=0.5$ and $z=1.5$, respectively.
  }
  \label{fig:MBH_Mstellar_Lbol_z_MLR}
\end{figure}

\begin{figure*}[ht!]
\gridline{\fig{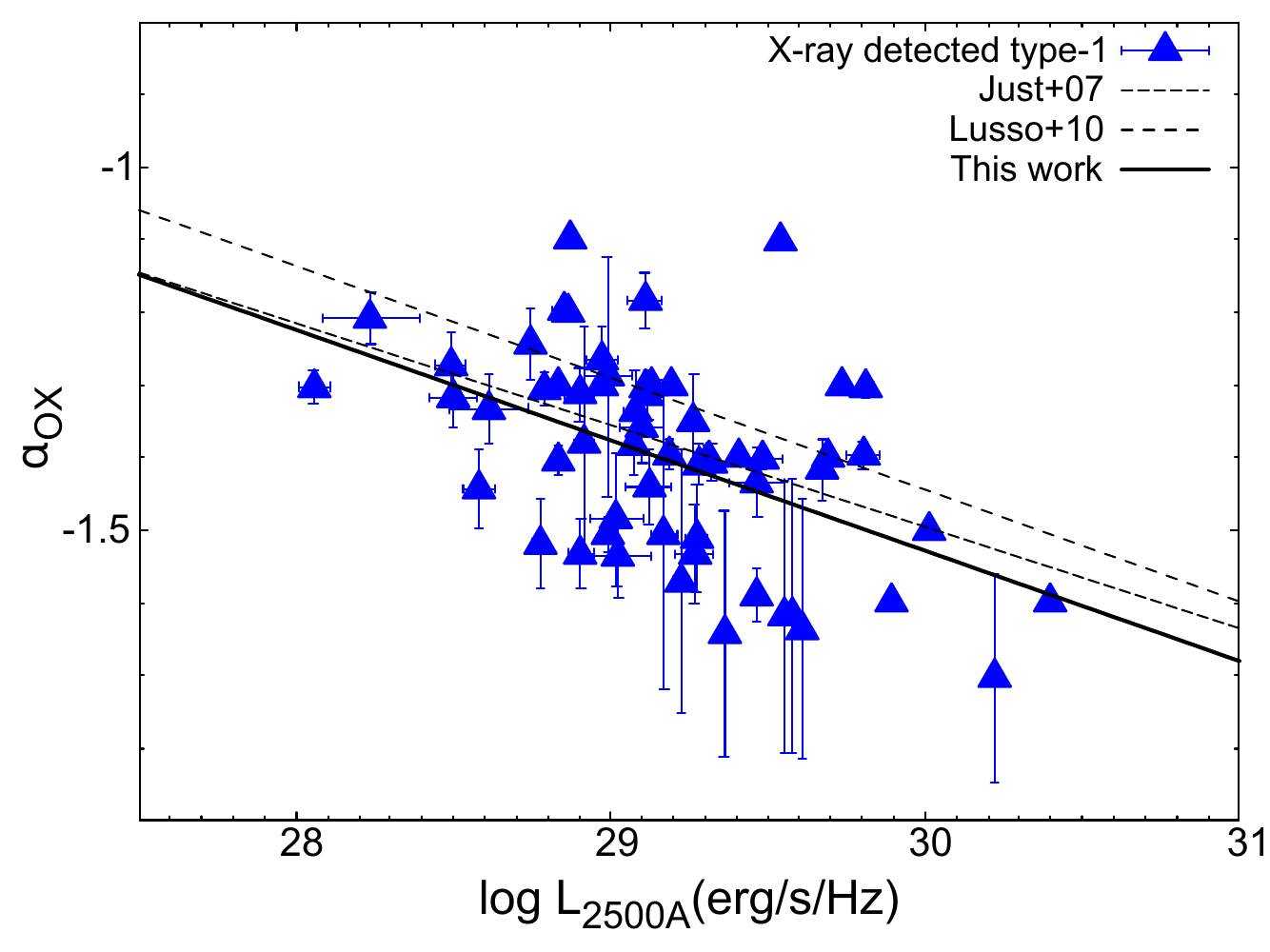}{0.45\textwidth}{(a)}
          \fig{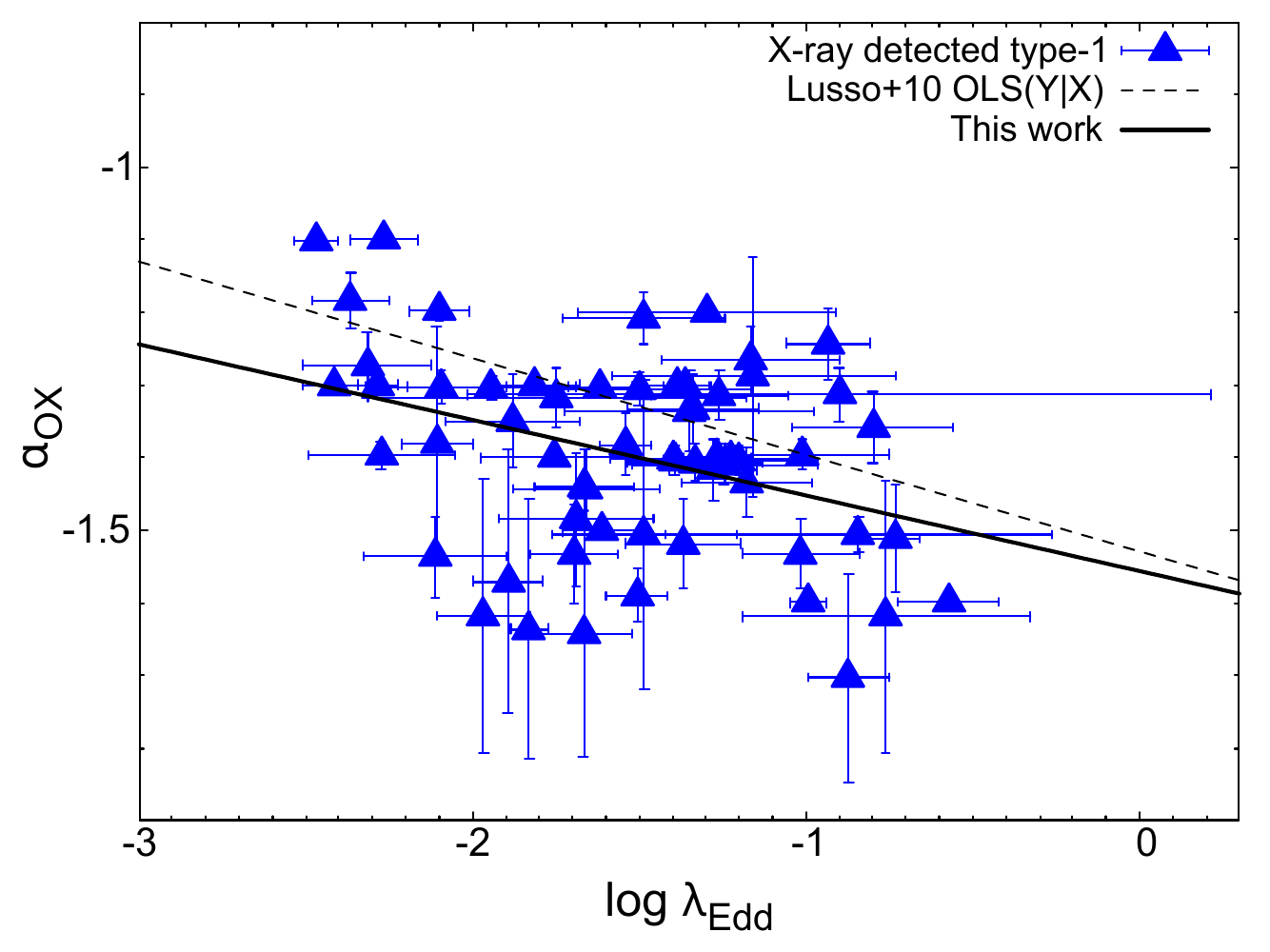}{0.45\textwidth}{(b)}
          }
\caption{$\alpha_{\rm ox}$ plotted against (a) $L_{2500\AA}$\ [erg s$^{-1}$ Hz$^{-1}$] and (b) $\lambda_{\rm Edd}$. The black solid and dashed lines in (a) represent the best-fit linear regression in Equation \ref{eq:alphaox_L2500A} and the relation of $\alpha_{\rm OX}$-$L_{2500\AA}$\ by \cite{Lusso10} and by \cite{Just07}, respectively. The black solid and dashed line in (b) shows the best-fit linear regression in Equation \ref{eq:alphaox_lambdaedd} and the $\alpha_{\rm OX}$-$\lambda_{\rm Edd}$ relation by \cite{Lusso10}, respectively.}
\label{fig:alphaox}
\end{figure*}

It is a key question which is the more important parameter, $z$
or $L_{\rm bol}$, that primarily determines the mean $M_{\rm
  BH}/M_{\rm stellar}$ ratio. Generally, it is difficult to separate the dependences because of the inevitable coupling between luminosity and redshift in a single flux-limited sample.
The combination of multiple surveys with different depths and widths is useful to better constrain it by expanding the coverage
in the luminosity versus redshift plane.

Figure \ref{fig:MBH_Mstellar_Lbol_z_MLR} plots $\log (M_{\rm BH}/M_{\rm stellar})$ against $\log L_{\rm
  bol}$, color-coded by redshift, for our type-1 AGN sample at $z=0.2-0.8$
(Stripe
82) and type-1 AGNs at $z\sim1.4$ in the Subaru/XMM-Newton Deep Survey
(SXDS) region \citep{Setoguchi21}. The averaged values of $(\log M_{\rm
  BH}/M_{\rm stellar})$ in different luminosity bins for the two samples
are plotted. This shows that $\log (M_{\rm BH}/M_{\rm stellar})$
increases with $\log L_{\rm bol}$, as already reported by
\citet{Setoguchi21} for the SXDS sample, whereas its redshift
dependence is weaker.
To confirm the above result, we perform a multiple linear-regression analysis among $\log (M_{\rm BH}/M_{\rm stellar})$, $z$, and $\log
L_{\rm bol}$ for the combined (Stripe 82 + SXDS) sample by utilizing
the python module {\tt statsmodels} \citep{seabold2010statsmodels}.
We obtain the following relation: 
\begin{eqnarray}
\log (M_{\rm BH}/M_{\rm stellar}) & = & (0.15\pm0.16)z \nonumber \\
 & + & (0.46\pm0.13)\log L_{\rm bol} \nonumber \\
 & + & (-23.15\pm5.61).
 \label{eq:MLR}
\end{eqnarray}
The best-fit lines in equation~\ref{eq:MLR} at two $z$ values ($z=0.5$ and 1.5) are plotted in Figure
\ref{fig:MBH_Mstellar_Lbol_z_MLR}. 
As \cite{Maji22} pointed out, to evaluate which variable the function has a stronger dependence, 
one must perform ``standardization'' of each parameter, that is,
subtracting the mean value and dividing the difference by the standard
deviation.
Then, we obtain the coefficients of $0.11\pm0.11$ and $0.41\pm0.11$
for {\it standardized} values of $z$ and $\log L_{\rm bol}$, respectively.
The larger coefficient in the latter term indicates that the
relation between $\log (M_{\rm BH}/M_{\rm stellar})$ and $\log L_{\rm
  bol}$ is the primary one.
The positive
luminosity dependence of $\log (M_{\rm BH}/M_{\rm stellar})$ may be a natural consequence that objects with larger $\log (M_{\rm BH}/M_{\rm stellar})$ ratios tend to show larger luminosities at
a given stellar mass when the $\log (M_{\rm BH}/M_{\rm stellar})$ distribution has an intrinsic scatter.
In fact, \cite{Li21b} performed detailed simulations and
found that the observed $M_{\rm BH}$-$M_{\rm stellar}$ ratios were 
biased toward higher values at higher redshifts in a 
flux-limited sample (i.e., luminosity-limited sample at a given
redshift) due to this effect.
Our result suggests that 
$\log (M_{\rm BH}/M_{\rm stellar})$ at given $\log L_{\rm bol}$
little evolves with redshift, supporting the conclusion by \cite{Li21b}.
Caution must be taken to interpret the above best-fit relation, however, because selection biases are complicated.
The weak redshift dependence is likely affected by the sample selection bias in a flux-limited sample (i.e., luminosity-limited sample at a given redshift). In fact, \cite{Li21b} performed detailed simulations and found that the observed $M_{\rm BH}$-$M_{\rm stellar}$ ratios are biased toward higher values at higher redshifts by assuming the ERDF by \cite{Schulze15}.
\ifnum0=1
The caution also must be taken that the connection between observed $(M_{\rm BH}/M_{\rm stellar})$-$L_{\rm bol}$ relation and selection biases is complicated. Although our analysis using a multiple linear regression implies that the redshift dependence on $(M_{\rm BH}/M_{\rm stellar})$ is weaker than the $L_{bol}$ dependence, magnitude limits can build up the $(M_{\rm BH}/M_{\rm stellar})$-$L_{\rm bol}$ relation because magnitude limits correspond to different $L_{\rm bol}$ limits at a different redshift.
\cite{Li21b} reported that the "observed" $M_{\rm BH}$-$M_{\rm stellar}$ relations are biased high at higher redshift assuming the ERDF by \cite{Schulze15}.
It is important to consider the SMBH properties (e.g., ERDF) effect on the $(M_{\rm BH}/M_{\rm stellar})$-$L_{\rm bol}$ relation or a sample detection limit because a $M_{\rm BH}$ limit changes along with $\lambda_{\rm Edd}$ at a given $L_{\rm bol}$ limit.
However, as noted above, determining the intrinsic ERDF by accounting for any selection biases is not a goal of our study. Our sample is limited and it is difficult to take all biases into account.
We hope that this work serves as a base for future studies on the connection among the $M_{\rm BH}-M_{\rm stellar}$ relation, selection biases, and SMBH-galaxy properties.
\fi

To summarize, we infer that the difference in the mean $\log (M_{\rm
BH}/M_{\rm stellar})$ values between AGNs with $\log L_{\rm bol}
\lesssim 45$ at $z \lesssim 1.2$ and those with $\log L_{\rm bol}
\gtrsim 45$ at $z \gtrsim 1.2$ can be explained solely by a common
luminosity dependence that does not evolve with redshift. It is
noteworthy in Table \ref{table:massratio_summary} that the majority of X-ray selected type-1 AGNs
at $z=0.2-0.8$ (this work), $z=0.5-1.1$ \citep{Schramm13}, and
$z=1.2-1.7$ \citep{Ding20}
\footnote{
\cite{Schawinski12} found similar results that dust-obscured galaxies (DOGs) at $z\sim2$ tend to have disk-dominant morphologies (i.e., not major mergers).}
have disk-like morphologies, although a
minor but significant fraction may have bulge-like ones. All these
results suggest that, at least for the majority of type-1 AGNs, there
are no distinct differences between the low luminosity, low redshift
AGNs and high luminosity, high redshift ones.
This provides no evidence for two distinct channels for SMBH growth to explain the
down-sizing behavior and seems to be more consistent with
theoretical models that consider common AGN triggering mechanisms over
a wide redshift range 
\citep[e.g.,][]{Shirakata19}.

We would like to make caveats here, however, that the above tentative conclusion is based purely on X-ray selected type-1 AGNs, and does not
include absorbed AGNs, a dominant X-ray AGN population at the low
luminosity range. In addition, there are mid-infrared selected, high
luminosity AGN populations that are not considered in our study, such as 
``reddened type-1 quasars'' \citep{Glikman18}. These
AGNs are known to be relatively X-ray weak (\citealt{Ricci17,Goulding18,Toba_19a}) and some of them (obscured ones)
show evidence for the merger channel of SMBH growth
(e.g., \citealt{Treister12,Donley18}).
To reach firm conclusions on the trigger mechanisms
of whole AGNs, it is crucial to investigate the nature of these
populations that are not included in this work.

\ifnum0=1
We confirmed that the $M_{\rm{BH}}-M_{\rm{stellar}}$ distribution is
close to that of our parent sample (see the full sample contours in
Figure 1 of \citealt{Li21b}). \cite{Matsuoka15} presented that the
$M_{\rm{BH}}-M_{\rm{stellar}}$ ratio of SDSS type-1 quasars located at
$z<1$ is also consistent with the local $M_{\rm{BH}}/M_{\rm{bulge}}$
one but flat $M_{\rm{BH}}-M_{\rm{stellar}}$ relation.
\cite{Matsuoka15} and \cite{Shen15} explained that the selection bias
in high-z luminosity-threshold quasars might cause this flattened
relation.
\fi

\ifnum0=1
\cite{Sun15} investigated 69 X-ray selected type-1 AGNs at z=0-2 with $43.8<\log\ L_{\rm bol}<47.2$ and concluded that no significant dependence of redshift on $M_{\rm{BH}}/M_{\rm{stellar}}$ is found.
If it is also true for our sample, the galactic disk mass included in $M_{stellar}$ is reallocated to grow up the bulge mass and to construct the $M_{\rm{BH}}-M_{\rm{bulge}}$ relation seen in $z\sim0$ with keeping the $M_{\rm{BH}}-M_{\rm{stellar}}$ ratio (e.g.,\citealt{Schramm13,Sun15,Ding20,Li21b}).
\cite{Reines15} reported that Seyfert 1 objects with $41.5<L_{\rm bol}<44.4$ and $\log\ M_{\rm BH}\lesssim8.0$ at $z<0.055$ tend to have smaller $M_{\rm{BH}}/M_{\rm{stellar}}$ ratios than the $M_{\rm{BH}}-M_{\rm{bulge}}$ ratio by \cite{Kormendy}.
The X-ray selected type-1 objects of \cite{Sun15} with $\log\  L_{\rm bol}<45$ also have the small $M_{\rm{BH}}-M_{\rm{bulge}}$ ratio which ranges from -4.08 to -2.36.
If the $M_{\rm{BH}}-M_{\rm{bulge}}$ relation in the local universe is not assembled, our sample, especially with small $M_{\rm BH}$ ($<10^8\ M_\odot$), may build the small $M_{\rm{BH}}-M_{\rm{stellar}}$ ratio.
\cite{Setoguchi21} exploring the moderately luminous (log $L_{\rm bol} \sim$ 44.5-46.5 erg s$^{-1}$) X-ray selected type-1 AGNs at $z\sim1.4$ reached the similar results but a slightly larger $M_{\rm{BH}}-M_{\rm{stellar}}$ ratio (log $(M_{\rm BH}/M_{\rm stellar})$ = -2.2) than this work although the larger $M_{\rm{BH}}-M_{\rm{stellar}}$ ratio can be caused by the selection bias toward luminous AGNs (as proposed in \citealt{Setoguchi21}).
If the population of objects in \cite{Setoguchi21} and that in this work are the same, (that is, if the host galaxies of luminous type-1 AGNs in \cite{Setoguchi21} are disk-dominant galaxies,) the secular star formation and/or the quench of the SMBH growth from $z=1.4$ to $z=0.5$ is needed to decrease the $M_{\rm BH}$-$M_{stellar}$ ratio and $L_{\rm bol}$ without experiencing major mergers.
If luminous type-1 AGNs in \cite{Setoguchi21} are hosted in bulge-dominant galaxies, it is possible that they experienced a different evolution scenario from our sample (e.g., major mergers) and keeping the $M_{\rm BH}-M_{\rm{stellar}}$ relation which can be considered as $M_{\rm BH}-M_{\rm{bulge}}$ one. In fact, elliptical galaxies tend to have massive SMBHs in the local universe \citep{Kormendy}.
We have to take care that stellar mass cut criteria which exclude smallest $M_{\rm stellar}$ ($<10^{9.3}M_\odot\ (z=0.2-0.4),\ 10^{9.7}M_\odot\ (z=0.4-0.6),\ {\rm and}\ 10^{10.3}M_\odot\ (z=0.6-0.8)$) and largest $M_{\rm stellar}$ objects ($>10^{11.5}$) with uncertain mass estimates were set to improve the mass completeness
(see Section 5 of \citealt{Li21a}).
\fi

\begin{figure*}[ht!]
\gridline{\fig{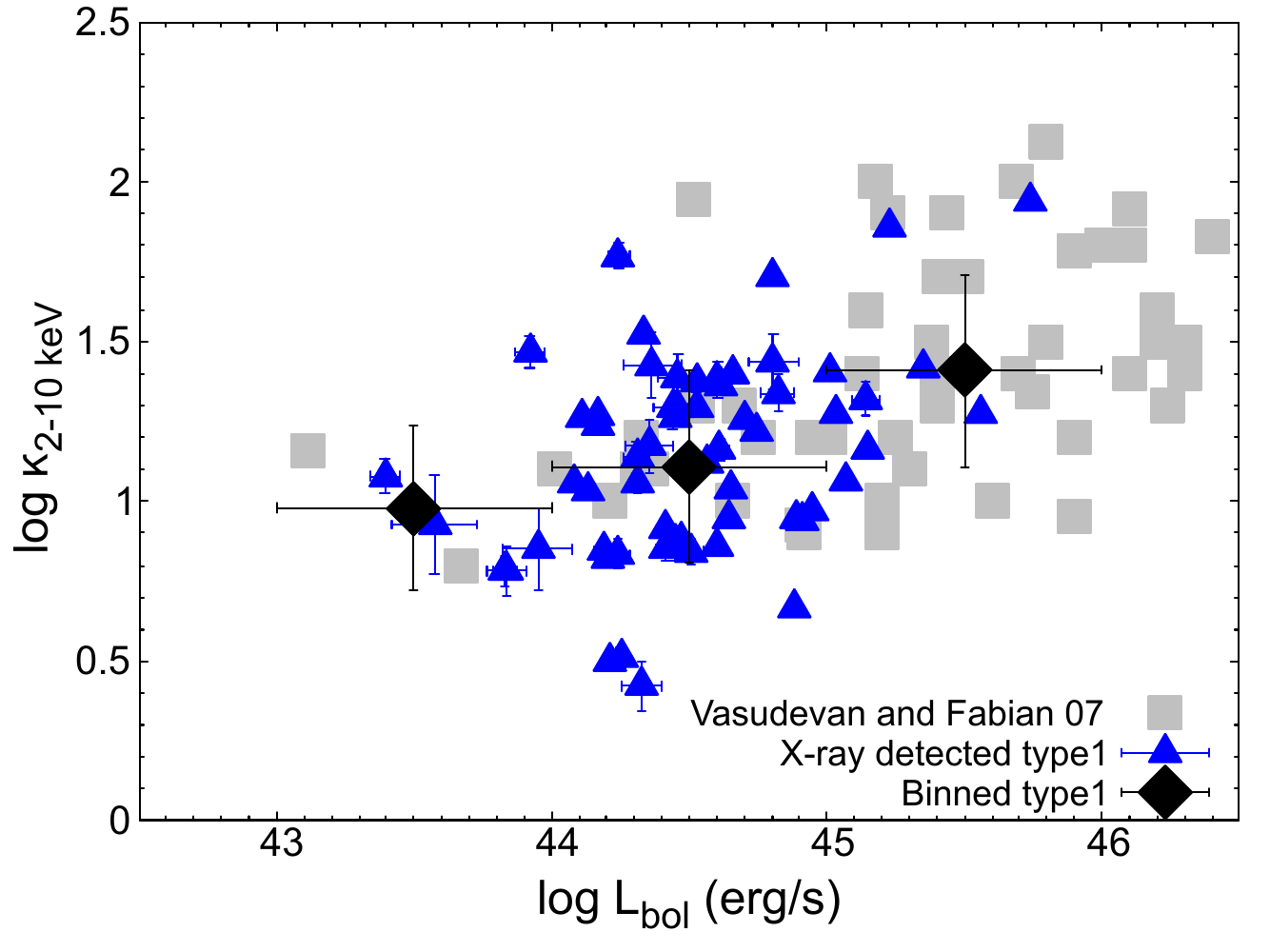}{0.45\textwidth}{(a)}
\fig{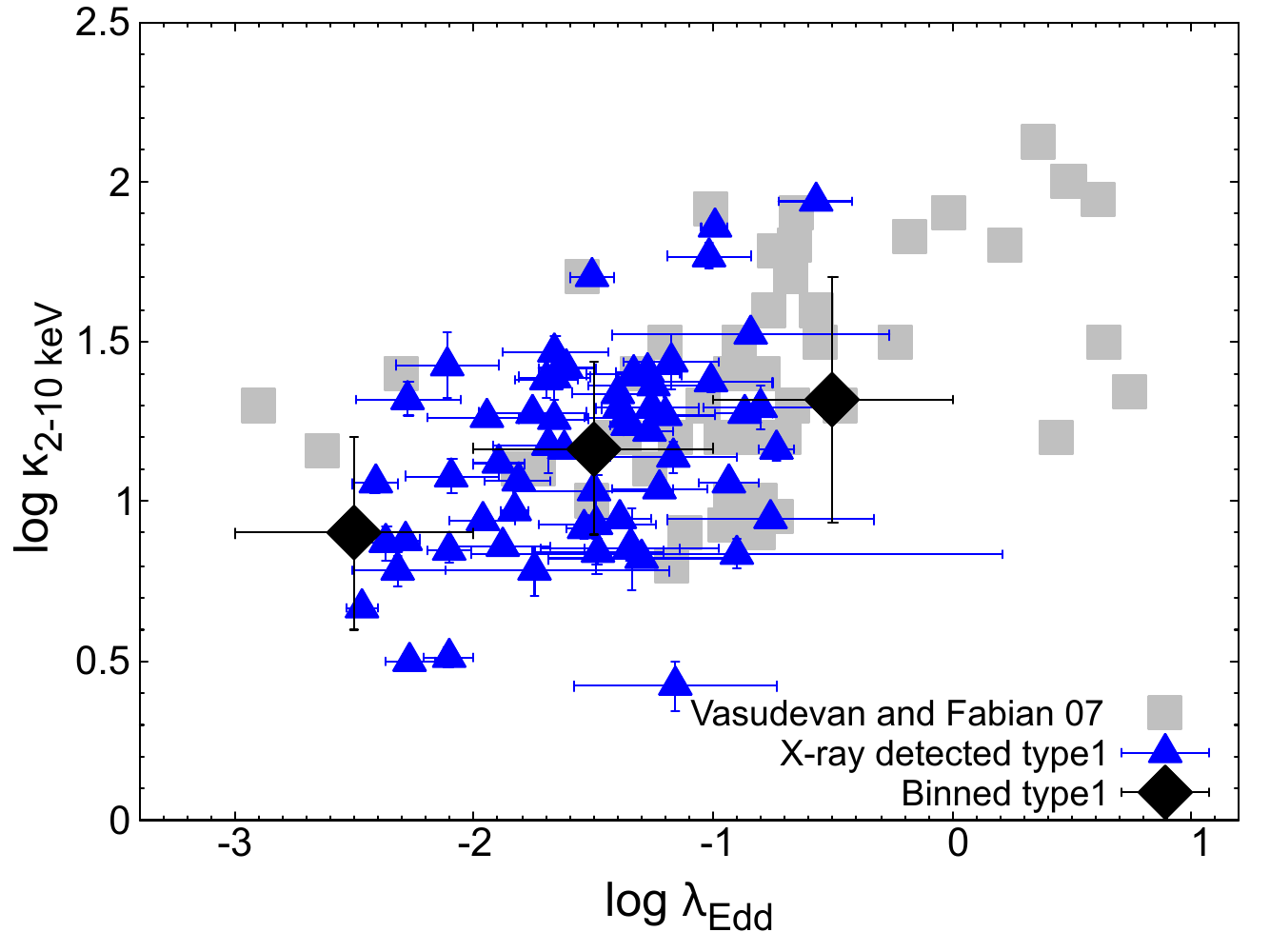}{0.45\textwidth}{(b)}}
\caption{$\kappa_{2-10}$ versus (a) $L_{\rm bol}$ or (b) $\lambda_{\rm Edd}$. The black diamonds display binned results. 
  The vertical position and the error bar show a mean value and a standard deviation of $\kappa_{2-10}$ in each bin, respectively. The grey squares represent AGNs from \cite{Vasudevan07}.} 
\label{fig:kappa}
\end{figure*}

\subsection{$\alpha_{\rm OX}$ versus $L_{2500\AA}$\ or $\lambda_{\rm Edd}$}

Many studies investigated the relationship between $\alpha_{\rm OX}$ and $L_{2500\AA}$\ or $\lambda_{\rm Edd}$ (e.g., \citealt{Vignali03,Strateva05,Steffen06,Shemmer06,Just07,Lusso10,Jin12,Martocchia17,Chiaraluce18}).
Using our sample, we plot the $\alpha_{\rm
  OX}$-$L_{2500\AA}$ 
\footnote{
 Here $L_{2500\AA}$ is the
intrinsic AGN luminosity at the rest-frame 2500\AA\ corrected for
extinction,
based on our best-fit AGN SED model
viewed at an inclination angle of 30\arcdeg.
}
and $\alpha_{\rm OX}$-$\lambda_{\rm Edd}$
relations in Figure \ref{fig:alphaox}. Their
correlation coefficients are found to be $r=-0.48\pm0.12$
($\alpha_{\rm OX}$-$L_{2500\AA}$) and $r=-0.58\pm0.12$ ($\alpha_{\rm
  OX}$-$\lambda_{\rm Edd}$), indicating significant anti-correlations
in both relations. 
We obtain linear regression forms between $\alpha_{\rm OX}$ and
$L_{2500\AA}$ or $\lambda_{\rm Edd}$ as:
\begin{eqnarray}
\alpha_{\rm OX} & = & (-0.15\pm0.04)\log\ (L_{2500\AA})+(3.03\pm1.03) \nonumber\\
 \label{eq:alphaox_L2500A}
\end{eqnarray}
and
\begin{eqnarray}
\alpha_{\rm OX} & = & (-0.10\pm0.04)\log\ \lambda_{\rm Edd}+(-1.56\pm0.06). \nonumber\\
 \label{eq:alphaox_lambdaedd}
\end{eqnarray}
These lines are shown in Figure \ref{fig:alphaox} together with
those derived by \cite{Just07} and \cite{Lusso10}.
The full sample of \cite{Just07} is composed of luminous quasars
(log $L_{2500\AA}\gtrsim 32$) and the samples from
\cite{Steffen06} and from \cite{Shemmer06}, covering a range of log
$L_{2500\AA}$ = 27.7--32.5 out to $z=4.5$.  \cite{Lusso10} selected
545 type-1 AGNs at $z=0.04-4.25$ with log {$L_{2500\AA}=25.7-31.4$}
detected in the XMM-COSMOS survey. 
As noticed, our results are well consistent with the results by \cite{Just07}.
We note that the mean value of $\alpha_{\rm OX}$ of our sample is
by $\sim 0.05$ smaller than the best-fit relations by \cite{Lusso10} at
the same $L_{2500\AA}$\ or $\lambda_{\rm Edd}$ ranges.
We infer that the differences are attributable to sample selection effects;
the sample of \cite{Lusso10} contains many AGNs at $z>1$ for which the detection limit of X-ray luminosity is higher than ours, and hence may
miss X-ray faint AGNs, whereas our sample may miss optically/UV faint
AGNs because of the magnitude limits in the SDSS. The scatter in
$\alpha_{\rm OX}$ around the best-fit relation is found to be $\pm0.35$ 
which is consistent with previous studies (e.g.,
\citealt{Just07,Lusso10}).

\ifnum0=1
To check the scatter in $\alpha_{\rm OX}$, we define
$\Delta\alpha_{\rm OX}=\alpha_{\rm OX}-\overline{\alpha_{\rm OX}}$, where $\overline{\alpha_{\rm OX}}$ is our regression line shown above.
We find that $\Delta\alpha_{\rm
  OX}$ of our sample ranges from $-0.30$ to $+0.27$ with a mean value of
$-0.08$. These results are in line with previous studies (e.g.,
\citealt{Just07,Lusso10}).
\fi

\ifnum0=1
We note that the mean value of $\alpha_{\rm OX}$ of our
sample is $-1.40$
which is similar to that of \cite{Lusso10} ($\alpha_{\rm OX}=-1.37\pm0.01$) but smaller
than that of the luminous quasars in \cite{Just07} ($\alpha_{\rm OX}=-1.80\pm0.02$). This is
because the luminous quasars in \cite{Just07} covers higher UV luminosities (log $L_{2500\AA}$ = 31.9--32.5) than our sample (log $L_{2500\AA}$ = 28.1--30.4) and the \cite{Lusso10} sample (log $L_{2500\AA}$ = 25.7--31.4).
\fi

\subsubsection{Bolometric Correction Factor $\kappa_{2-10}$ versus $\lambda_{\rm Edd}$ or $L_{\rm bol}$}

We also calculate the 2--10 keV-to-bolometric luminosity correction
factor defined as $\kappa_{2-10}\equiv L_{\rm bol}/L_{\rm 2-10 keV}$.
\ifnum0=1
We note that the SKIRTOR model implemented in the current version of
CIGALE assumes a fixed spectral shape in the optical to UV bands as AGN disk emission, although in reality, it should be complex, and is likely a function of black hole mass and mass accretion rate (e.g., \citealt{Kubota18}).
\fi
The relations between $\kappa_{2-10}$ and $\lambda_{\rm Edd}$ or
$L_{\rm bol}$ are shown in Figure \ref{fig:kappa}. We find the trend
that $\kappa_{2-10}$ increases with $\lambda_{\rm Edd}$ or $L_{\rm
  bol}$ in our objects. We also plot the mean values of $\kappa_{2-10}$ in (a) $L_{\rm bol}$ ($43<\log\ L_{\rm bol}<44$, $44<\log\ L_{\rm
  bol}<45$, and $45<\log\ L_{\rm bol}<46$) or (b)
$\log \lambda_{\rm Edd}$
($\log\ \lambda_{\rm Edd}<-2$, $-2<\log\ \lambda_{\rm Edd}<-1$, and
$-1<\log\ \lambda_{\rm Edd}$),
confirming the trend.
\cite{Vasudevan07} found a similar relation in type-1
AGNs with $z<0.7$ at $-3 < \log \lambda_{\rm Edd} < 1$ but the small number of objects satisfied $\log \lambda_{\rm Edd} <-2$. Our result confirms that the correlation continues to even lower Eddington rates of 
$\log \lambda_{\rm Edd} <-2$.
This may be consistent with the disk truncation scenario as suggested in changing look AGNs (e.g., \citealt{Noda18}). 
Thus, $\kappa_{2-10}$ may be used as a beacon of the mass accretion rates normalized by the black hole mass,
as previously pointed out (e.g., \citealt{Vasudevan07,Lusso10}),
over a wider range of $\log \lambda_{\rm Edd}$.

\subsection{Comparison of X-ray detected type-1 and type-2 AGNs}
\label{sec:control}

\ifnum0=1
\begin{figure}[ht!]
\plotone{AGNratio_Lbol_morp_spec.pdf}
\caption{The relative fraction of X-ray detected type-2 AGNs plotted against $L_{\rm bol}$. The horizontal position and bar show the center and the width of the bin, respectively. The vertical bar corresponds to the relative error.}
\label{fig:AGNratio}
\end{figure}
\fi

We compare the AGN and host properties between the X-ray detected type-1 and type-2 AGN samples. 
In Figure \ref{fig:histogram}, we display the histograms of
$M_{\rm stellar}$, $L_{\rm bol}$, and $L_{\rm bol}/M_{\rm stellar}$
for the type-2 AGN sample (red) to be compared with those for the type-2 AGN
sample (blue). 
The median, mean, and
standard deviation of these parameters are summarized in Table
\ref{table:output_parameter}.
Figure~\ref{fig:coevolution_parameters}(c) shows the $L_{\rm bol}$
versus $M_{\rm stellar}$ plot. As evident in this figure,
type-2 AGNs are more abundant in the low $L_{\rm bol}$ region.
A Kolmogorov-Smirnov (KS) test for the $L_{\rm bol}$ distribution gives
$p=3.2\times10^{-4}$, indicating a significant difference between the
type-1 and type-2 AGN samples.
In other words, the observed fraction of type-2 AGNs in the total AGNs
decreases with bolometric luminosity. To derive the intrinsic type-2
AGN fraction as a function of $L_{\rm bol}$ is beyond the scope of this paper because our samples are not complete and complex selection biases must be corrected.

\ifnum0=1
To confirm this trend in a different way, we calculated the fraction of
  type-2 AGNs among the whole AGNs, which are 
  $0.760\pm0.180\ (\log L_{\rm bol}<45)$ and
  $0.574\pm0.186\ (\log L_{\rm bol}>45)$.
  We note that these type-2 fractions must be taken with caveats because our samples are not complete to pure X-ray selection (e.g., SDSS spectroscopic data are
required) and complex bias correction must be applied to derive the
intrinsic fraction. As seen in Figure
  \ref{fig:L5100_MBHestimate} (a), the host galaxy contribution to
  optical spectra is large in low-luminosity AGNs. It is possible that
  lower-luminosity type-1 AGNs may be classified as type-2 AGNs in the
  SDSS catalog because broad lines in lower-luminosity type-1 AGNs may
  not be observed due to low signal-to-noise ratios. At the lowest
luminosity range, where the type-2 fraction is close to unity, a
contribution from a ``true type-2'' population (i.e., those having no
broad line regions intrinsically), may not be negligible.
Hence it is difficult to accurately evaluate these
  selection biases and estimating the intrinsic obscured AGN fraction
  is beyond the scope of this paper.
\fi

To test the AGN unified scheme, it is important to check whether the
host steller-mass distributions of type-1 and type-2 AGNs with given
AGN properties are the same or not. Recalling the fact that $L_{\rm
  bol}$ is more correlated with $M_{\rm BH}$ than with $\lambda_{\rm
  Edd}$ in type-1 AGNs (see
Section~\ref{subsubsec:MBH_Lbol_Mstellar}), it is not fair to directly
compare the $M_{\rm stellar}$ distributions using the {\it whole}
type-1 and type-2 AGN samples, given that type-2 AGNs tend to have
lower luminosities (i.e., biased toward lower SMBH masses). Thus, we
divide the samples by $L_{\rm bol}$, those with $\log L_{\rm
  bol}>44.5$ and with $\log L_{\rm bol}<44.5$ for each AGN type. We
find that the $M_{\rm stellar}$ distribution of type-2 AGNs is similar
to that of type-1 AGNs in both luminosity regions; KS tests yield
$p=5.1\times10^{-2}$ ($\log L_{\rm bol}>44.5$) and $p=2.9\times10^{-1}$ ($\log L_{\rm
  bol}<44.5$).

Thus, type-1 and type-2 AGNs with common luminosity (likely SMBH mass)
ranges share similar host properties in terms of stellar mass. It is
quite important because it helps to justify the use of host stellar
mass, in place of black hole mass, to roughly estimate Eddington
ratios commonly in type-1 and type-2 AGNs, as done in many
studies. Our results support the ``luminosity dependent'' unified
scheme
(\citealt{Ueda03,LaFranca04,Simpson05,Hasinger08,Ueda14,Toba14}), or
probably ``Eddington-ratio dependent'' unified scheme \citep{Ricci17};
that is, type-1 and type-2 AGNs belong to the same population of host
galaxies with obscuring AGN tori whose covering fraction decreases
with luminosity (or Eddington ratio).
\ifnum0=1
This figure suggests that type-2 AGNs have a large fraction in the lower $L_{\rm bol}$ even though both ends of bins have large relative errors. Previous studies investigated the dependence of the type-1 (or type-2) AGN fraction on $L_{\rm bol}$.  For example, \cite{Simpson05}
reported that type-1 AGN fraction increases slowly with AGN
luminosities. \cite{Ueda14} found that the absorbed AGN fraction and
AGN luminosities are anti-correlated in both lower and higher
redshift.  Our result is consistent with these studies.
These findings align with the AGN unification paradigm, indicating that our results are consistent with this framework.
\fi

\section{Conclusion}\label{sec:conclusion}

In this study, we conducted a comprehensive analysis of X-ray detected
AGNs with multiwavelength counterparts in the Stripe 82 region at
$z=0.2-0.8$. The sample consisted of 60 type-1 AGNs and 137 type-2
AGNs, spanning an X-ray luminosity range of $\log L_{\rm X} = 41.6-44.7$
We utilized the latest CIGALE code, which 
includes dusty polar components, to carry out the analysis.
To obtain accurate parameters of both the AGN and host galaxy in
type-1 AGNs, we utilized the image-decomposed optical SEDs obtained by
\cite{Li21a} based on Subaru Hyper Suprime-Cam (HSC) images. We
estimated reliable black hole masses using the host and total SEDs by 
subtracting the host galaxy contribution in the continuum luminosity at
5100\AA.

Our conclusions are summarized as follows.
\begin{enumerate}
    \item The mean value of log $(M_{\rm BH}$/$M_{\rm stellar})$ in our type-1 AGN sample is found to be $-2.7\pm0.5$, is similar to the local mass ratio between black holes and bulges. 

    \item Performing a multi-linear regression analysis to a combined
  sample of type-1 AGNs in the Stripe 82 region and the SXDF
  \citep{Setoguchi21}, we find that $\log (M_{\rm BH}/M_{\rm stellar})$
  depends primarily on the AGN-luminosity, not on redshift. The 
  offset in $\log (M_{\rm BH}/M_{\rm stellar})$ between our type-1 AGN
  sample and more luminous ($\log L_{\rm bol} > 45$) type-1 AGNs at
  $z\sim1.5$ can be attributed to its luminosity dependence.
    
    \item
  We find anti-correlations between the UV-to-X-ray slope ($\alpha_{\rm OX}$) and AGN luminosity or Eddington ratio, which are consistent with previous studies. We confirm the trend that the X-ray-to-bolometric correction factor ($\kappa_{2-10}$) increases with Eddington ratio by covering a range of
      $\log \lambda_{\rm Edd} <-2$.

    \item Our type-1 and type-2 AGNs with the same luminosity ranges share similar distributions of $M_{\rm stellar}$, whereas type-2 AGNs exhibit smaller $L_{\rm bol}$ on average than type-1 AGNs. This supports the luminosity (or Eddington ratio) dependent unified scheme.

\end{enumerate}


\begin{acknowledgments}
  We are very grateful for the anonymous referee for a careful reading
  of the manuscript and providing many
  useful comments, which helped us greatly improve the clarity of the paper.
This work was financially supported by the Grant-in-Aid for Scientific
Research grant Nos. 21J23187 (K.S.), 20H01946 (Y.U.), 19K14759 and 22H01266 (Y.T.).
\end{acknowledgments}
%

\vspace{5mm}


\software{astropy \citep{2013A&A...558A..33A,2018AJ....156..123A}, {\tt CIGALE} v2022.0 \citep{Yang22}, XSPEC \citep{Arnaud96}
          }



\appendix

\section{Comparison of X-ray detected and undetected type-1 AGNs}

\begin{figure*}[ht!]
\gridline{\fig{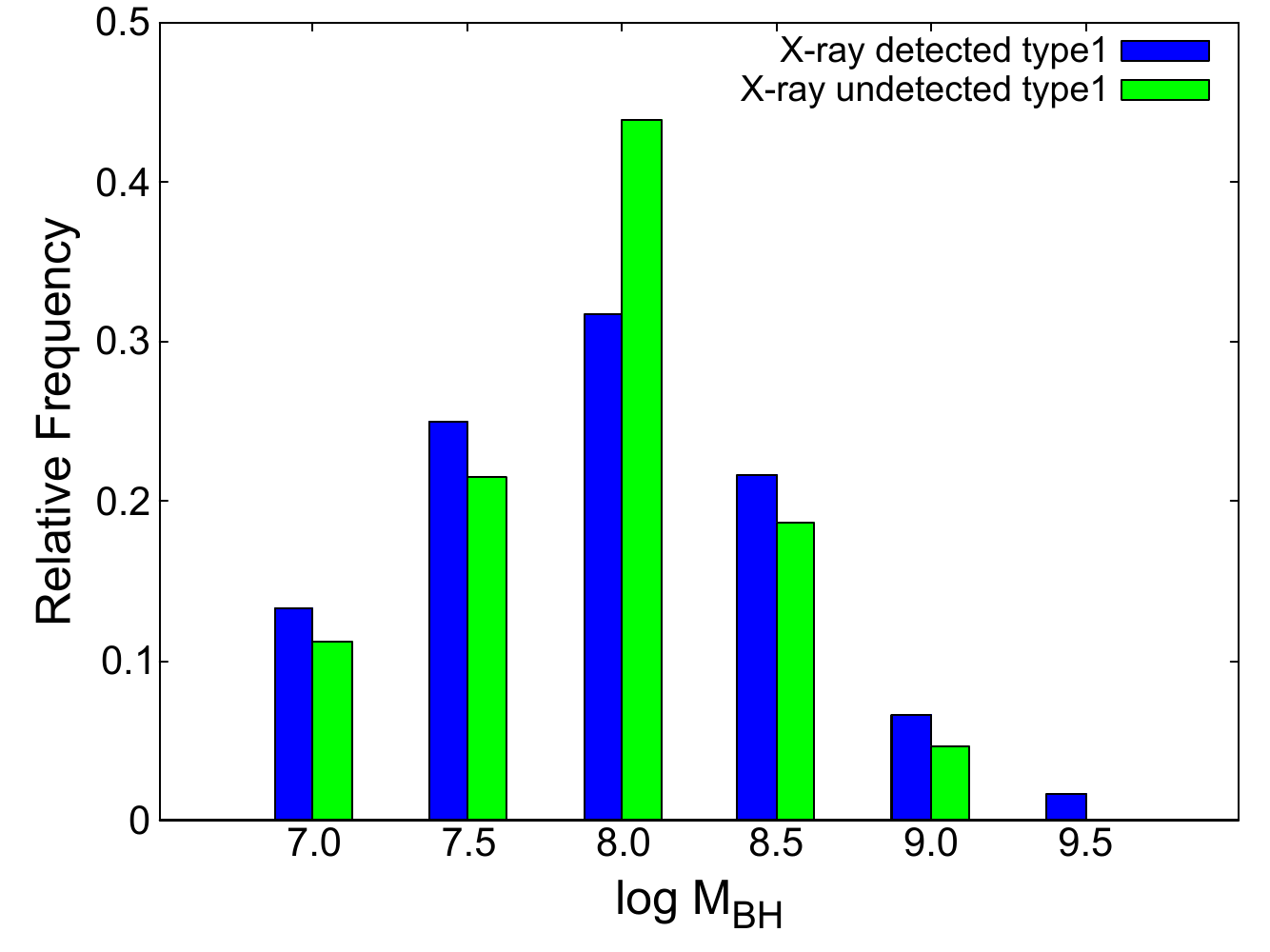}{0.3\textwidth}{(a)}\fig{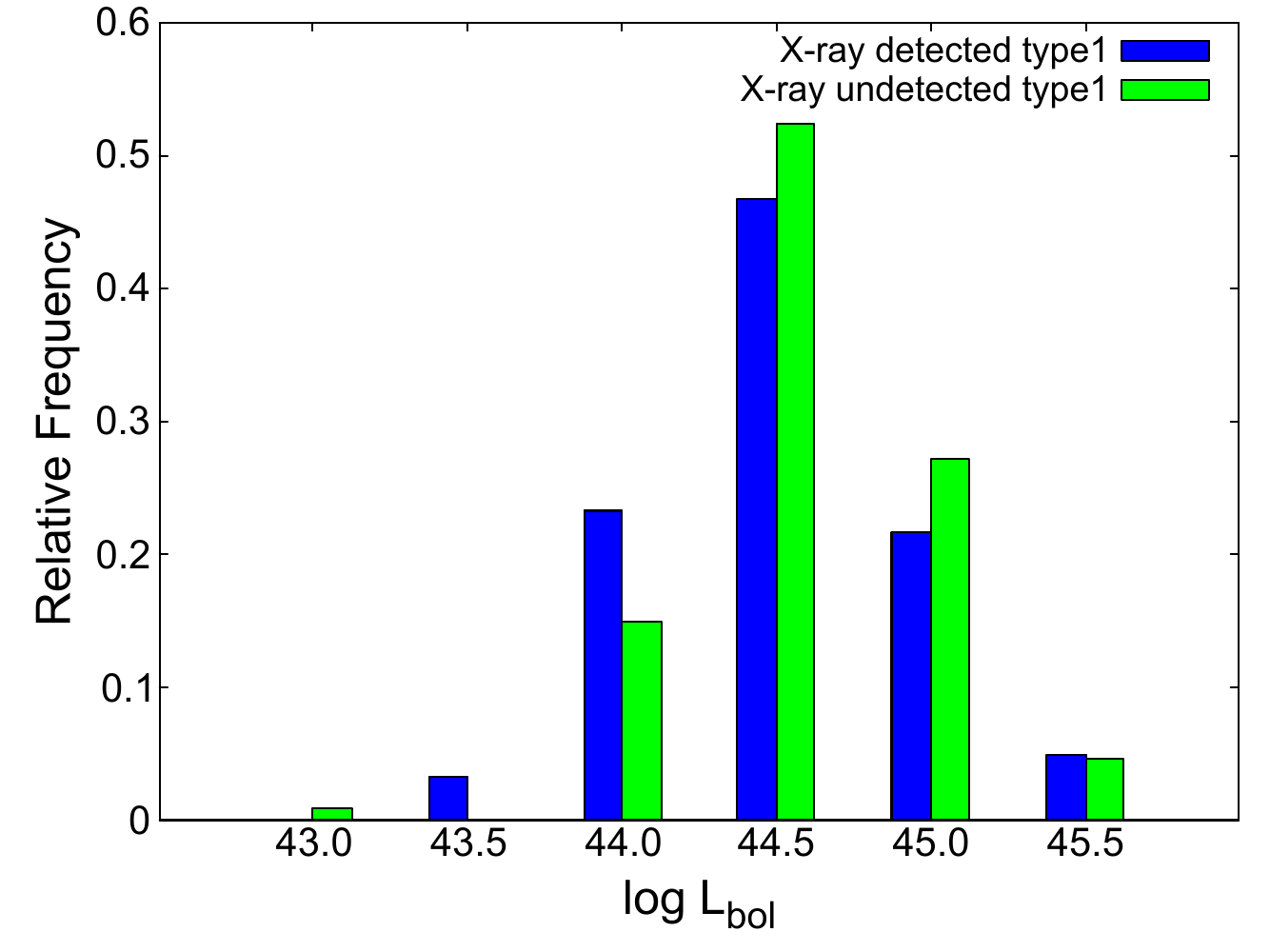}{0.33\textwidth}{(b)}
\fig{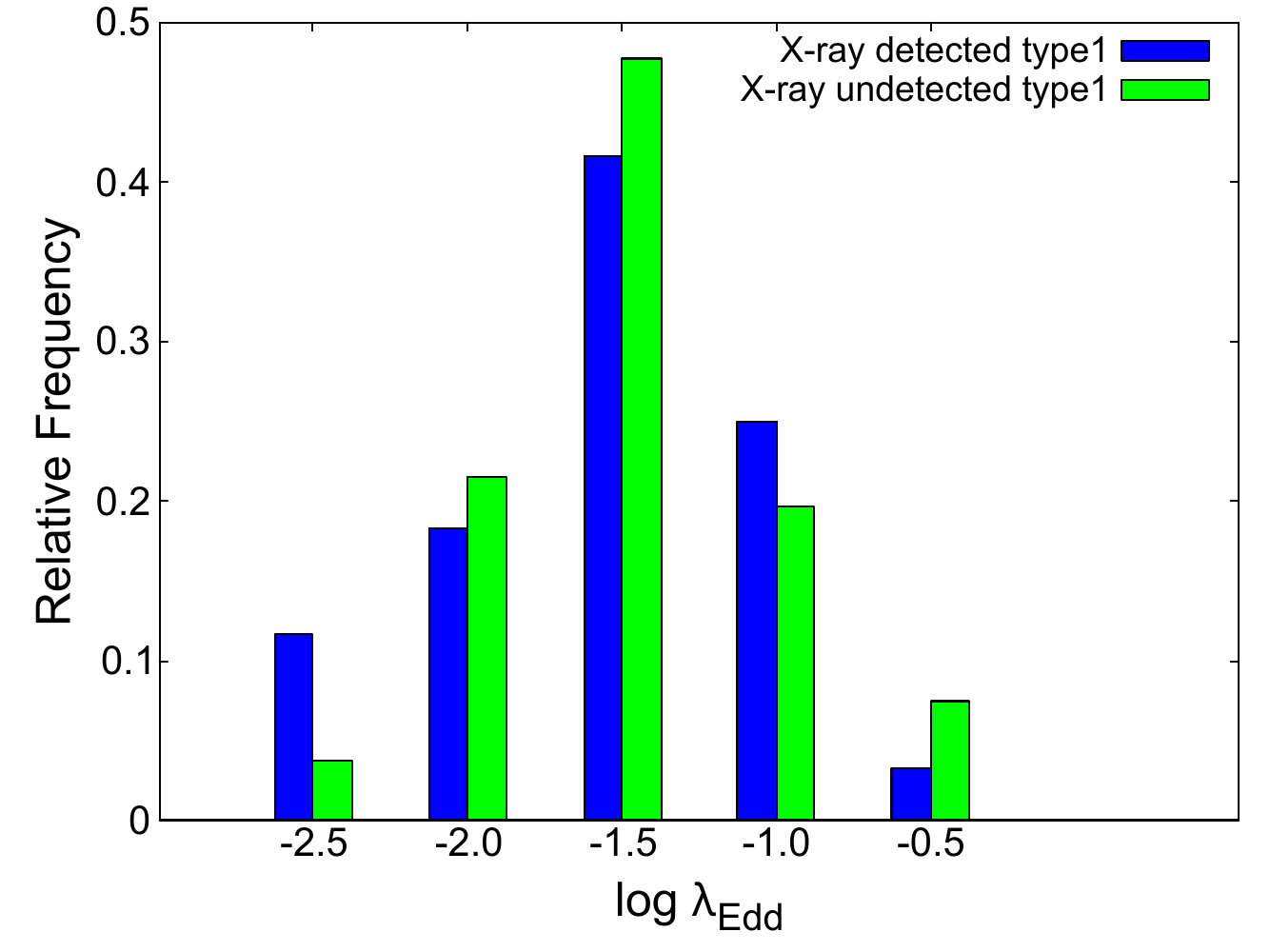}{0.3\textwidth}{(c)}}
\gridline{\fig{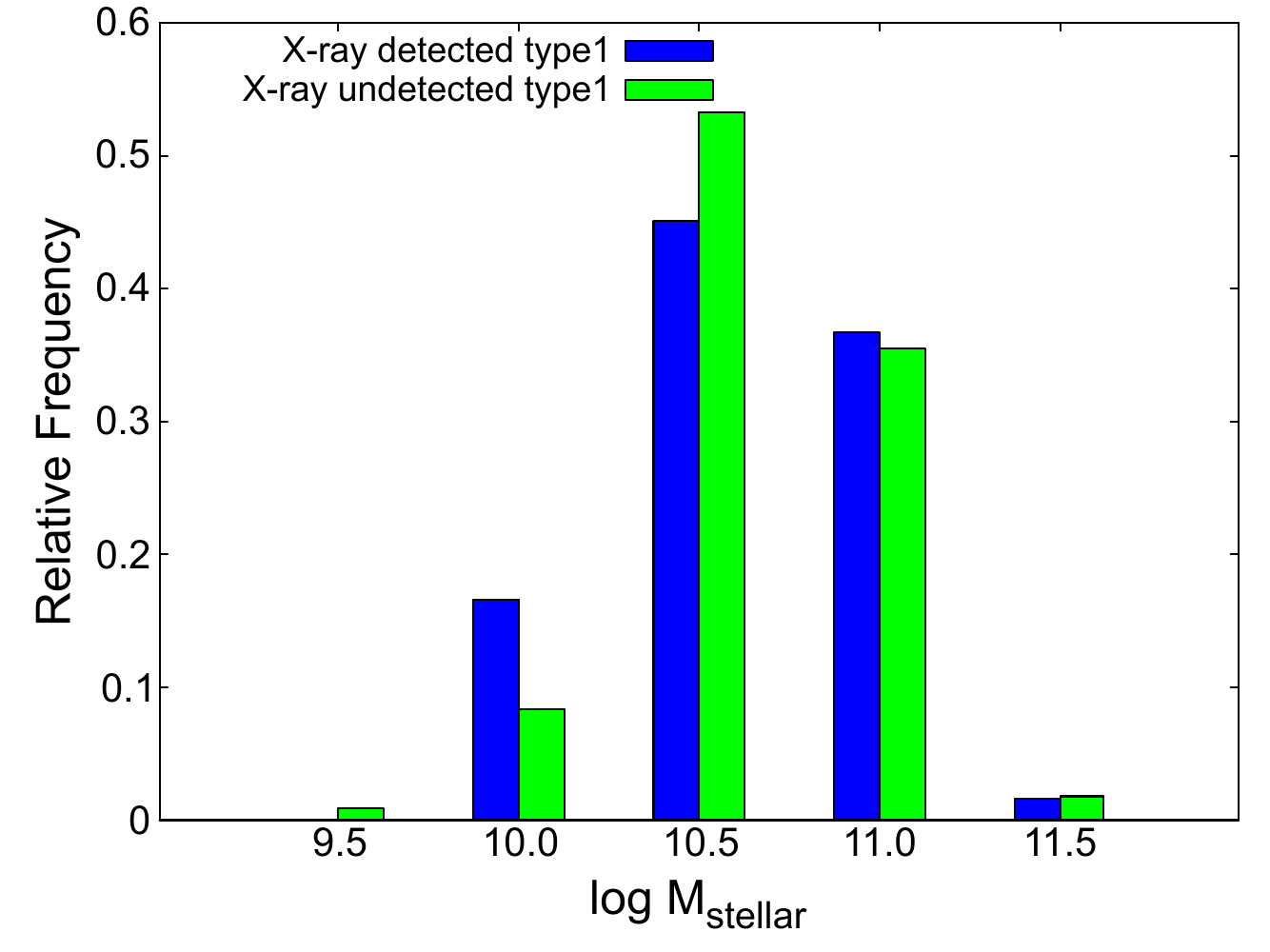}{0.3\textwidth}{(d)}\fig{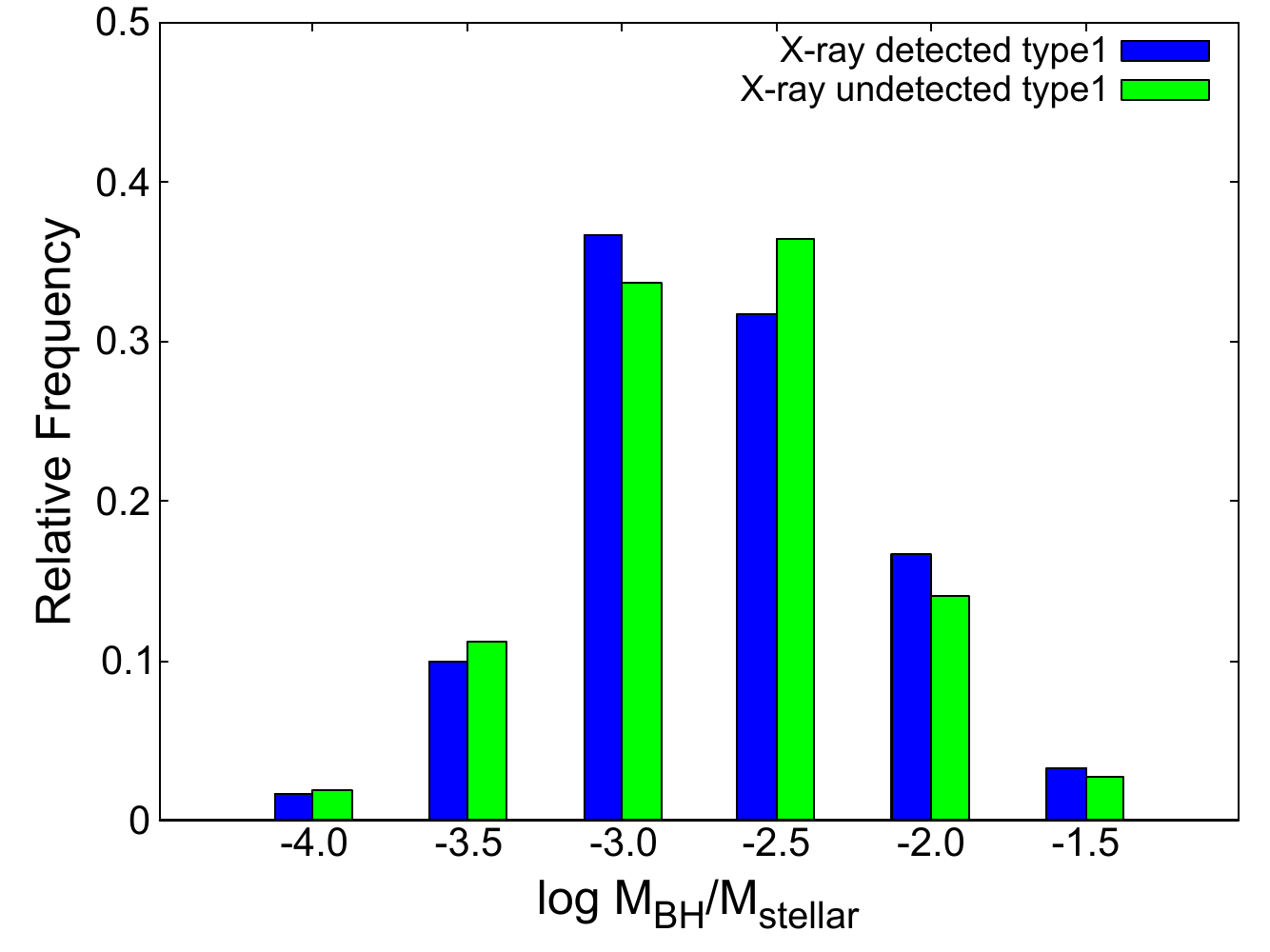}{0.3\textwidth}{(e)}
\fig{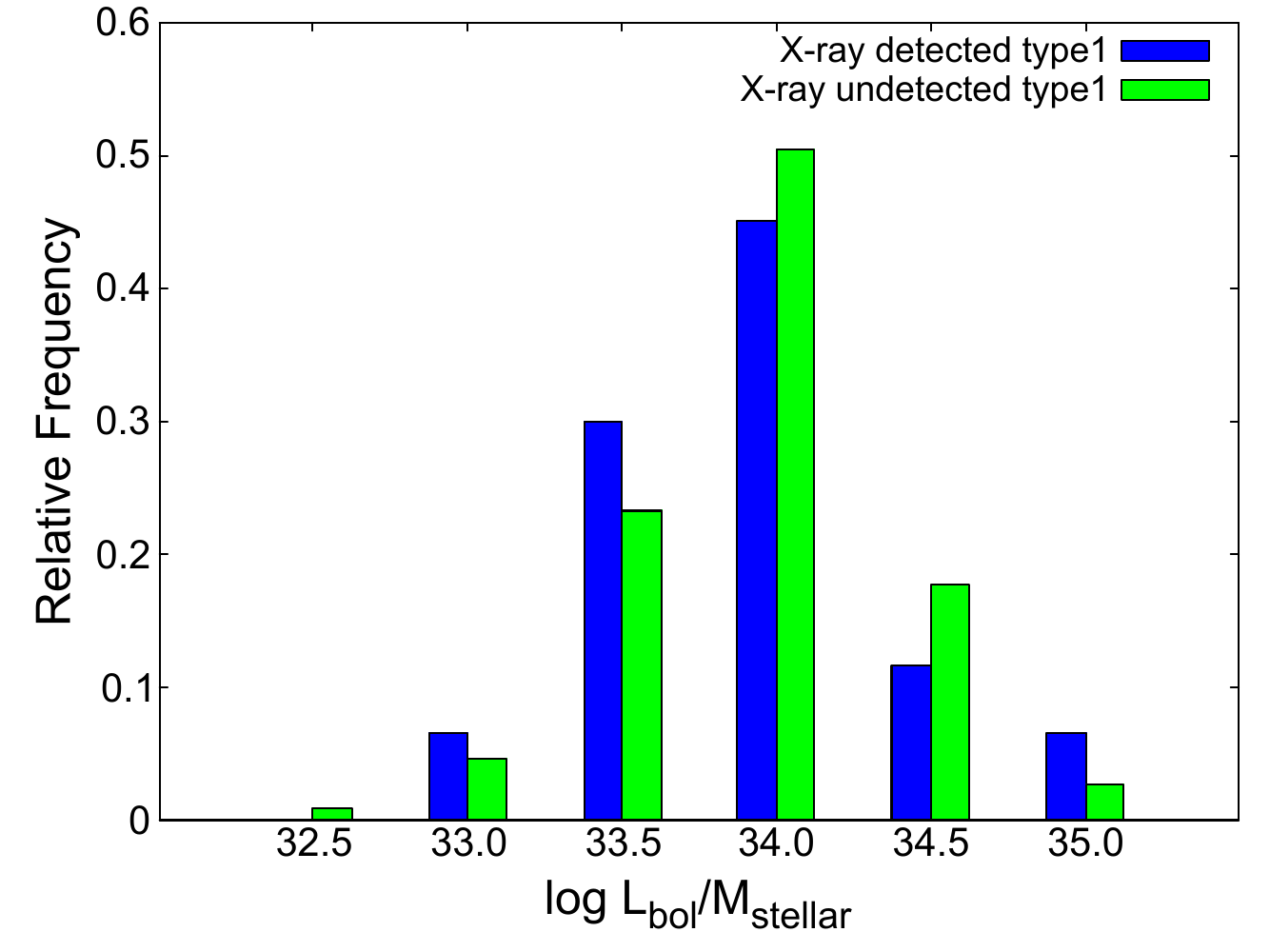}{0.3\textwidth}{(f)}}
\gridline{\fig{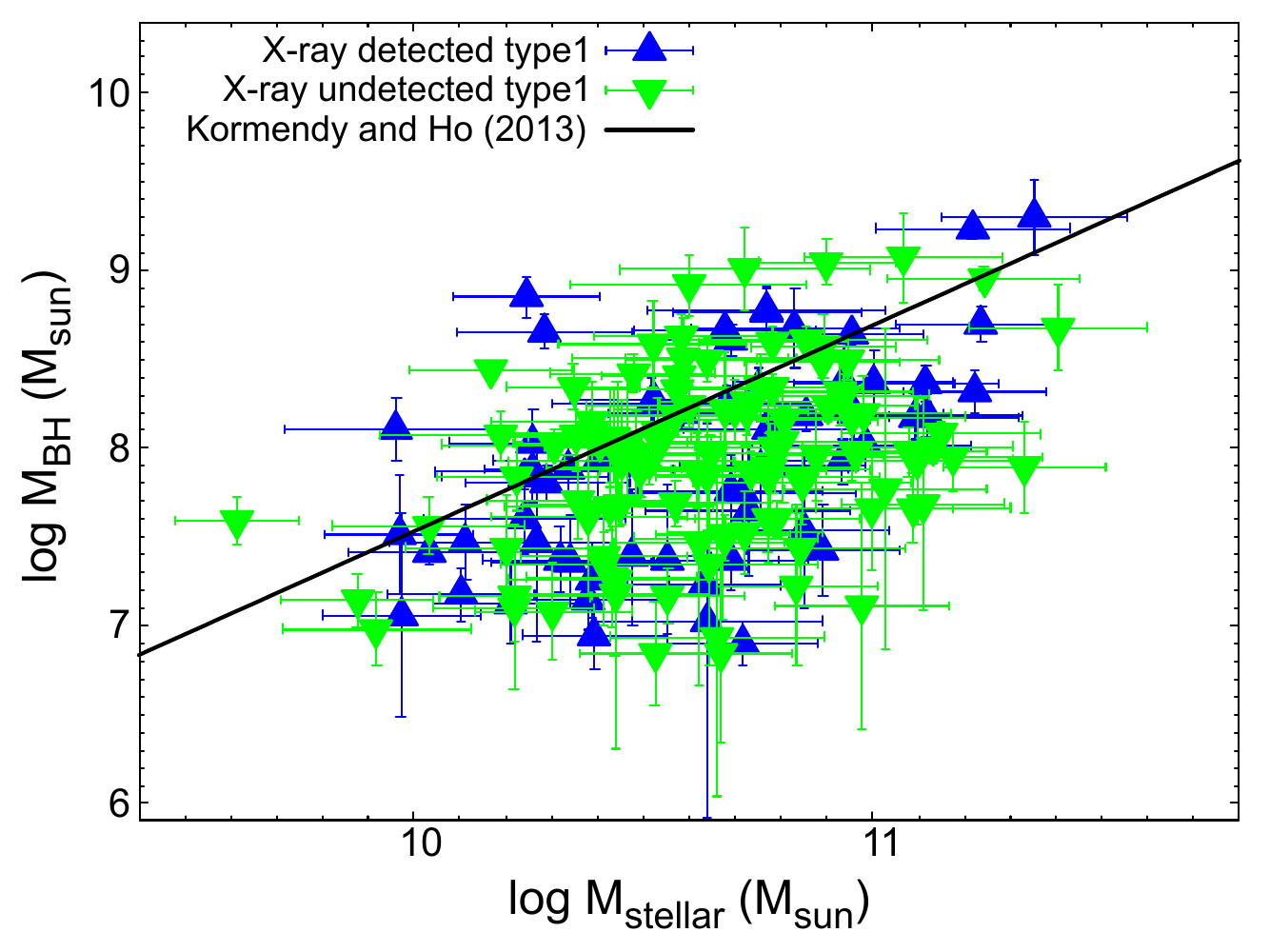}{0.3\textwidth}{(g)}}
\caption{Comparison of the histograms of (a) log $M_{\rm BH}\ [M_\odot]$, (b) log $L_{\rm bol}$ (erg s$^{-1}$), (c) log $\lambda_{\rm Edd}$, (d) log $M_{\rm stellar}\ [M_\odot]$, (e) log $(M_{\rm BH}/M_{\rm stellar})$, and (f) log $(L_{\rm bol}/M_{\rm stellar})$ in X-ray detected (blue) and undetected (green) objects. (g) The $M_{\rm BH}$-$M_{\rm stellar}$ relation. The green inverse triangles represent X-ray undetected type-1 AGNs. The black line corresponds to the local $M_{\rm BH}$-$M_{\rm bulge}$ relation by \cite{Kormendy}.}\label{fig:xray_undetected}
\end{figure*}

Here we investigate whether X-ray detected and undetected type-1 AGNs show any systematic differences in the host and AGN parameters. 
As the comparison sample, we selected X-ray undetected type-1 AGNs at $z=0.2-0.8$ from the quasar host catalog \citep{Li21a} in the Stripe 82 region. These objects have counterparts in the SDSS DR14Q catalog within 1\arcsec.
The selection criteria are as follows:
\begin{enumerate}
    \item No counterpart in the Stripe 82X catalog within 5\arcsec 
    \item No detection in the 3XMM-DR7 catalog 
\end{enumerate}
The SED analysis of the X-ray undetected type-1 AGNs was carried out in
the same way as for the X-ray detected type-1 AGNs except that
the X-ray model is not included. We utilize 15
photometries (or their upper limits) from radio to UV
(VLA, WISE, 2MASS or UKIDSS, Subaru, GALEX)
bands taken from the type-1 quasar multiwavelength catalogs
(\citealt{Paris18,Rakshit20}). The SEDs of 107 objects out of the 139 objects were well reproduced with reduced $\chi^2<10$. 

Figure~\ref{fig:xray_undetected} (a-f) compares the histograms of $M_{\rm
  BH}$, $L_{\rm bol}$, 
$\lambda_{\rm Edd}$, $M_{\rm stellar}$, $M_{\rm BH}/M_{\rm stellar}$,
$L_{\rm bol}$/$M_{\rm stellar}$, 
respectively, between the X-ray detected and undetected type-1 AGN samples.
By performing Kolmogorov-Smirnov (KS) tests, we find no significant
differences in these distributions between the two samples.
The $M_{\rm BH}$-$M_{\rm stellar}$ relation is displayed in Figure~\ref{fig:xray_undetected} (g), again showing no significant
difference between the two samples.
These results suggest that X-ray detection does not cause any biases in selecting type-1 AGNs. Most probably, X-ray detection or nondetection is determined by time variability, which is faster in X-ray bands (through Comptonizing corona) than in the optical band (accretion disk).


\bibliography{sample631}{}
\bibliographystyle{aasjournal}



\end{document}